\documentclass[aps,prc,superscriptaddress,twoside,twocolumn,nofootinbib,10pt,%
showpacs,floatfix]{revtex4-1}

\usepackage{amsmath,amssymb}
\usepackage{graphicx,bm}
\usepackage{slashed}
\usepackage{epstopdf}
\usepackage{ulem} 
\usepackage[usenames]{color}
\usepackage{subfigure}
\usepackage{float}

\newcommand{\Slash}[1]{{\ooalign{\hfil/\hfil\crcr$#1$}}}

\allowdisplaybreaks

\begin{document}

\title{Appearance of novel modes of $\bar{D}$ mesons with negative velocity in the dual chiral density wave}

\author{Daiki Suenaga}
\email{suenaga@hken.phys.nagoya-u.ac.jp}
\affiliation{Department of Physics,  Nagoya University, Nagoya, 464-8602, Japan}

\author{Masayasu Harada}
\email{harada@hken.phys.nagoya-u.ac.jp}
\affiliation{Department of Physics,  Nagoya University, Nagoya, 464-8602, Japan}

\date{\today}

\newcommand\sect[1]{\emph{#1}---}
\begin{abstract}
We calculate dispersion relations for 
$\bar{D}\, (0^-)$, $\bar{D}^*\, (1^-)$, $\bar{D}_0^*\, (0^+)$ and $\bar{D}_1\, (1^+)$ mesons 
in the Dual Chiral Density Wave (DCDW) 
where the chiral symmetry is spontaneously broken by 
inhomogeneous chiral condensate. 
Employing the Bloch's theorem,
we show that 
all the 
modes have  
opposite  group velocity to the momentum in the low-momentum region and the energy is minimized at non-zero momentum.
Furthermore, the magnitude of momentum which realizes the minimum energy 
is equal to the wave number of the density wave.  
\end{abstract}
\maketitle

\section{Introduction}

The spontaneous chiral symmetry breaking (S$\chi$SB) is one of the most important features of the quantum chromodynamics (QCD) in the low-energy region.
In the vacuum the S$\chi$SB is driven by the non-vanishing chiral condensate, $\langle \bar{q} q \rangle \neq0$, which is expected to change at non-vanishing temperature and/or density.
Studying hadron properties in such an extreme environment will provide some clues to understand more about QCD.
Several literatures~\cite{Kunihiro:1993py,Schon:2000qy,Basar:2008im,Nickel:2009wj,Abuki:2011pf,Buballa:2014tba,%
Sadzikowski:2000ap,Nakano:2004cd,Heinz:2013hza} 
showed that, in the high density region, there emerges a phase in which the chiral symmetry is broken by the inhomogeneous chiral condensate, i.e. the value of the $\langle\bar{q} q\rangle$ depends on the position such as in the dual chiral density wave (DCDW) or the chiral density wave (CDW). 
Recently, Ref.~\cite{Heinz:2013hza} showed that, using a hadronic effective model, the CDW 
is realized 
at the hadron level in the density region about $2.4$ times of the normal nuclear matter density.
Furthermore, using a model of the Skyrme-crystal type, Ref.~\cite{Harada:2015lma} pointed a similarity between the DCDW and the half-Skyrmion phase, where the chiral symmetry is broken by an inhomogeneous condensates, and that the phase transition occurs at about twice of the normal matter density. 
The inhomogeneous condensate in the density region about twice of the normal nuclear matter density, if exists, could be checked by the RISP experiment planned in Korea.  
So it is very interesting to study the changes of the hadron properties caused by such an inhomogeneous condensate.

In Refs.~\cite{Suenaga:2014dia,Suenaga:2014sga},
we proposed to use the heavy-light mesons such as  $\bar{c}q$ mesons composed of one anti-charm quark $\bar{c}$ and one light quark $q$ (up quark or down quark) as probes to explore the phase with an inhomogeneous condensate such as the 
DCDW.
There are several works which study the properties of the heavy-light mesons in the nuclear matter by using hadronic models based on, e.g., 
a
heavy meson effective model based on the heavy quark symmetry~\cite{Yasui:2012rw}, and a model with $SU(8)$ symmetry~\cite{Tolos:2009nn}.
In Ref.~\cite{Suenaga:2014dia}, we pointed out that the existence of spin-isospin correlation in nuclear matter causes the mixing among heavy-light mesons having different spins.
Then,
by using an effective hadron model based on the heavy quark symmetry and the chiral symmetry, it was shown that $D(J^P=0^-)$ and $D^*(1^-)$ mesons mix with each other in spin-isospin correlated nuclear matter such as in the DCDW.
In Ref.~\cite{Suenaga:2014sga}, the analysis was extended to include 
the $D_0^\ast(0^+)$ and $D_1(1^+)$ mesons 
as chiral partners to $D(0^-)$ and $D^*(1^-)$ mesons~\cite{heavy-partner}, and the mass spectra of the charmed mesons in the half-Skyrmion phase were studied. 

In this paper, we study the dispersion relations for $\bar{D}\, (0^-)$, $\bar{D}^*\, (1^-)$, $\bar{D}_0^*\, (0^+)$ and $\bar{D}_1\, (1^+)$, which we collectively call ``$\bar{D}$ mesons'', in the DCDW by 
regarding $(\bar{D},\bar{D}^*)=(0^-,1^-)$ and $(\bar{D}_0^*,\bar{D}_1)=(0^+,1^+)$ as the chiral
partners to each other.
It should be noted that we do not use 
``$D$ mesons''  but use ``$\bar{D}$ mesons ''
in the present work, since ``$D$ mesons'' include anti-light quarks which provide annihilation processes.
Then, we assume that the ``$\bar{D}$ mesons''
pick up the effect of nuclear matter only mediated by the exchange of mesons made of light quarks.
Here we introduce interactions of the ``$\bar{D}$ mesons'' with a matrix field $M=\sigma+i\tau^a\pi^a$  ($\tau^a$ is the Pauli matrix) including a scalar meson ($\sigma$) and pions ($\pi^a$), which transforms linearly under the chiral symmetry.
We replace the field $M$ with its classical configuration in the DCDW, and regard the configuration as a potential for the ``$\bar{D}$ mesons''.
Since the potential in the DCDW is periodic in a space direction, 
we employ the Bloch's theorem in order to obtain the dispersion relations for ``$\bar{D}$ mesons''. Our results show that 
all the modes have a group velocity opposite to the momentum  
in the low-momentum region.
Furthermore, the
minimum of the energy is realized when 
the magnitude of
the momentum is equal to the wave number of the density wave. 
  
 This paper is organized as follows. In Sec~\ref{sec:lagrangianwithpion}, we introduce an effective Lagrangian for the ``$\bar{D}$ mesons'' interacting with pions in the relativistic form.
  In Sec~\ref{sec:EOMs}, we 
derive the equation of motions of ``$\bar{D}$ mesons'' in the presence of the DCDW background in the matrix form. In Sec~\ref{sec:Dispersions}, we show the dispersion relations for ``$\bar{D}$ 
mesons''.
In Sec~\ref{sec:DiscussionAndSummary}, we give a summary and discussions.
Construction of a heavy meson effective model with the chiral partner structure is summarized in Appendix~\ref{sec:HMET}, and the Bloch's theorem is briefly reviewed  in Appendix~\ref{sec:BlochTheorem}. 
We show a derivation of the dispersion relation in Appendix~\ref{sec:ExtendedZone}. 


\section{$\bar{D}$ mesons Lagrangian interacting with pions}
\label{sec:lagrangianwithpion}

In the present analysis, we regard $\bar{D}_0^*$ meson with $J^P = 0^+$ and $\bar{D}_1$ meson with $J^P= 1^+$ as the chiral partners to $\bar{D}$ meson with $j^P=0^-$ and $\bar{D}^\ast$ meson with $J^P=1^-$, respectively.  Then, the mass difference between the partners comes from the effect of chiral symmetry breaking~\cite{heavy-partner}. 
On the other hand, 
$\bar{D}$ and $\bar{D}^*$ mesons, and $\bar{D}_0^*$ and $\bar{D}_1$ mesons are degenerated respectively due to the heavy quark symmetry in the limit of infinite charm quark mass. 
Using this idea, we can easily construct an effective Lagrangian of heavy-light mesons interacting with pions in the framework of the Heavy Meson Effective Theory (HMET) based on the $SU(2)_h$ heavy quark spin symmetry, the $SU(2)_L\times  SU(2)_R$ chiral symmetry and the parity invariance. We summarize main points of the HMET combined with the chiral partner structure relevant to the present analysis in Appendix~\ref{sec:HMET}. 
In Eq.~(\ref{LagrangianOfHMET}) we show an effective Lagrangian of ``$\bar{D}$ mesons'' 
written in terms of $H$ and $G$ doublets made of ($\bar{D}$, $\bar{D}^\ast$) and ($\bar{D}_0^\ast$, $\bar{D}_1$), respectively.
Based on the Lagrangian in the heavy quark limit, we construct the following Lagrangian in the relativistic form, which we use in the present analysis:~\footnote{%
We can reproduce the Lagrangian~(\ref{LagrangianOfHMET}) by 
substituting 
\begin{eqnarray}
\bar{D} = \frac{1}{\sqrt{m}}{\rm e}^{-imv\cdot x}\bar{D}_v \ ,
\nonumber
\end{eqnarray}
where $v$ is the velocity of $\bar{D}$ meson, 
and similarly for the $\bar{D}^\ast$, $\bar{D}_0^\ast$ and $\bar{D}_1$ mesons,
and neglecting $O(1/m)$ corrections.
}
\begin{widetext}
\begin{eqnarray}
{\cal L} &=&   \partial_{\mu}\bar{D}_0^*\partial^{\mu}\bar{D}_0^{* \dagger}-m^2\bar{D}_0^*\bar{D}_0^{*\dagger}-\partial_{\mu}\bar{D}_{1\nu}\partial^{\mu}\bar{D}_1^{\dagger\nu}+\partial_{\mu}\bar{D}_{1\nu}\partial^{\nu}\bar{D}_1^{\dagger\mu}+m^2\bar{D}_{1\mu}\bar{D}_1^{\dagger\mu} \nonumber\\
&& +\partial_{\mu}\bar{D}\partial^{\mu}\bar{D}^{\dagger}-m^2\bar{D}\bar{D}^{\dagger}-\partial_{\mu}\bar{D}^*_{\nu}\partial^{\mu}\bar{D}^{*\dagger\nu}+\partial_{\mu}\bar{D}^*_{\nu}\partial^{\nu}\bar{D}^{*\dagger\mu}+m^2\bar{D}^*_{\mu}\bar{D}^{*\dagger\mu} \nonumber\\
&& -\frac{1}{2}\frac{m\Delta_m}{f_{\pi}}[\bar{D}_0^*(M+M^{\dagger})\bar{D}_0^{*\dagger}-\bar{D}_{1\mu}(M+M^{\dagger})\bar{D}_1^{\dagger\mu}-\bar{D}(M+M^{\dagger})\bar{D}^{\dagger}+\bar{D}_{\mu}^*(M+M^{\dagger})\bar{D}^{*\mu\dagger}] \nonumber\\
 && -\frac{1}{2}\frac{m\Delta_m}{f_{\pi}}[\bar{D}_0^*(M-M^{\dagger})\bar{D}^{\dagger}-\bar{D}_{1\mu}(M-M^{\dagger})\bar{D}^{*\dagger\mu}-\bar{D}(M-M^{\dagger})\bar{D}_0^{*\dagger}+\bar{D}_{\mu}^{* }(M-M^{\dagger})\bar{D}_1^{\dagger\mu}]\nonumber\\ 
&& -\frac{g}{2}\frac{m}{f_{\pi}}[\bar{D}_1^{\mu}(\partial_{\mu}M^{\dagger}-\partial_{\mu}M)\bar{D}_0^{*\dagger}-\bar{D}_0^{*}(\partial_{\mu}M^{\dagger}-\partial_{\mu}M)\bar{D}_1^{\dagger\mu}-\frac{1}{m}\epsilon^{\mu\nu\rho\sigma}\bar{D}_{1\mu}(\partial_{\nu}M^{\dagger}-\partial_{\nu}M)i\partial_{\sigma}\bar{D}_{1\rho}^{\dagger}] \nonumber\\
&& +\frac{g}{2}\frac{m}{f_{\pi}}[\bar{D}^{*\mu}(\partial_{\mu}M^{\dagger}-\partial_{\mu}M)\bar{D}^{\dagger}-\bar{D}(\partial_{\mu}M^{\dagger}-\partial_{\mu}M)\bar{D}^{*\dagger\mu}-\frac{1}{m}\epsilon^{\mu\nu\rho\sigma}\bar{D}_{\mu}^*(\partial_{\nu}M^{\dagger}-\partial_{\nu}M)i\partial_{\sigma}\bar{D}_{\rho}^{*\dagger}] \nonumber\\
&& +\frac{g}{2}\frac{m}{f_{\pi}}[\bar{D}_1^{\mu}(\partial_{\mu}M^{\dagger}+\partial_{\mu}M)\bar{D}^{\dagger}+\bar{D}(\partial_{\mu}M^{\dagger}+\partial_{\mu}M)\bar{D}_1^{\dagger\mu}]\nonumber\\
& & -\frac{g}{2}\frac{m}{f_{\pi}}[\bar{D}_0^*(\partial_{\mu}M^{\dagger}+\partial_{\mu}M)\bar{D}^{*\dagger\mu}+\bar{D}^{*\mu}(\partial_{\mu}M^{\dagger}+\partial_{\mu}M)\bar{D}_0^{*\dagger}]\nonumber\\
& & -\frac{g}{2}\frac{1}{f_{\pi}}[\epsilon^{\mu\nu\rho\sigma}\bar{D}_{1\nu}(\partial_{\rho}M^{\dagger}+\partial_{\rho}M)i\partial_{\sigma}\bar{D}_{\mu}^{*\dagger}+\epsilon^{\mu\nu\rho\sigma}\bar{D}_{\mu}^*(\partial_{\rho}M^{\dagger}+\partial_{\rho}M)i\partial_{\sigma}\bar{D}_{1\nu}^{\dagger}] \ ,\label{StartingLagrangian}
\end{eqnarray}
\end{widetext}
where $M$ is the chiral field in the linear realization defined as $M=\sigma+i\pi^a\tau^a$ with $\sigma$ and $\pi^a$ being the scalar and pseudo-scalar fields, respectively, and $\tau^a$ being the Pauli matrix.
In Lagrangian~(\ref{StartingLagrangian}),
$m$ is the average mass of $G$ doublet and $H$ doublet, and $\Delta_m$ and $g$ are model parameters. 
In the vacuum the field $M$ has a vacuum expectation value (VEV) as $\langle M\rangle=\langle M^{\dagger}\rangle=f_{\pi}$, which generates 
a 
mass difference between two doublets.
In this paper, we fix the VEV from
 the pion decay constant $f_{\pi}=92.4$\,MeV.
The value of $|g|$ is estimated from the width of $\Gamma\left( D^{*+}\to D^0\pi^+ \right)$ experimentally as $|g| = 0.50$.~\footnote{
We use $\Gamma\left(D^{*+}\to D^0\pi^+\right) = \frac{|g|^2m^2\vert p_\pi \vert^3}{12\pi m_H^2}=56 \,\mbox{keV}$ together with $m_H = 1.97\,\mbox{GeV}$ and $m=2.19\,\mbox{GeV}$.
}
We determine the value of $\Delta_m$ from the difference of masses of $H$ and $G$ doublets as
\begin{eqnarray}
m_G^2 &=& m^2+m\Delta_m \ ,\nonumber\\
m^2_H &=& m^2-m\Delta_m  \ ,
\end{eqnarray}
where the masses of $G$ doublet and $H$ doublet are estimated by the spin average as
\begin{eqnarray}
m_G &=& \frac{m_{\bar{D}_0^*}+3m_{\bar{D}_1}}{4} = 2.40\, \mbox{GeV}\ ,\nonumber\\
m_H &=& \frac{m_{\bar{D}}+3m_{\bar{D}^*}}{4} = 1.97\,\mbox{GeV}\ .
\end{eqnarray}
Using these values, $m$ and $\Delta_m$ are estimated as
\begin{eqnarray}
m &=& \frac{m_G+m_H}{2} = 2.19\ {\rm GeV}\ ,\nonumber\\
\Delta_m &=& m_G-m_H = 430\ {\rm MeV}\ .
\end{eqnarray}

We should note that each of $\bar{D}^\ast$ and $\bar{D}_1$ fields describes three spin-states in 
addition to the multiplicity in the iso-spin space, 
since these mesons are massive spin one particles. However, the Lagrangian~(\ref{StartingLagrangian}) is written in terms of $\bar{D}^{*\mu}$ and $\bar{D}_1^{\mu}$, 
which means that each of these fields contains four degrees of freedom: $\bar{D}^{*\mu=0}\cdots \bar{D}^{*\mu=3}$ and $\bar{D}_1^{\mu=0}\cdots \bar{D}_1^{\mu=3}$.
So we have to eliminate one degree of freedom by a certain constraint.
In the vacuum, 
the equations of motion obtained from the Lagrangian in Eq.~(\ref{StartingLagrangian}) leads to the Lorentz condition $\partial_\mu \bar{D}^{\ast \mu} = 0$ and $\partial_\mu \bar{D}_1^\mu = 0$.
In the present case, on the other hand, 
we are interested in the dispersion relations for ``$\bar{D}$ mesons'' in the nuclear medium, i.e., the DCDW.
This will be done by solving the equations of motion of ``$\bar{D}$ mesons'' which are corrected by the medium effects.  
Then the Lorenz condition will generally be modified by the medium effects.
Here we assume that three dimensional momentum of ``$\bar{D}$ mesons'' and derivative of the  pion field $U$ are of order $\Lambda_{\rm QCD}$, and that 
the energy of ``$\bar{D}$ mesons'' is of the order of the heavy quark mass $m$. 
Then, the Loranz conditions are modified as
\begin{eqnarray}
\partial_{\mu}\bar{D}^{*\mu} &=& O(\Lambda_{\rm QCD}^2\bar{D}_\alpha/m) \nonumber\\
\partial_{\mu}\bar{D}_1^{\mu} &=& O(\Lambda_{\rm QCD}^2\bar{D}_\alpha/m)
\end{eqnarray}
where $\bar{D}_{\alpha}$ indicates $\bar{D},\bar{D}_0^*$ mesons and space components of $\bar{D}^*$ and $\bar{D}_1$ mesons.
 From these constraints, we can notice that the time components of $\bar{D}^*$ and $\bar{D}_1$ fields 
are expressed by the other fields as 
$\Lambda_{\rm QCD}D_\alpha/m$. 
Then inserting these constraints into the Lagrangian~(\ref{StartingLagrangian}) and eliminating time component of $\bar{D}^*$ and $\bar{D}_1$ mesons, we can get the Lagrangian in terms of $\bar{D},\bar{D}^{*i}, \bar{D}_0^*$ and $\bar{D}_1^i$ as follows:
\begin{widetext}
\begin{eqnarray}
{\cal L} &=&   \partial_{\mu}\bar{D}_0^*\partial^{\mu}\bar{D}_0^{* \dagger}-m^2\bar{D}_0^*\bar{D}_0^{*\dagger}-\partial_{\mu}\bar{D}_{1i}\partial^{\mu}\bar{D}_1^{\dagger i}+m^2\bar{D}_{1i}\bar{D}_1^{\dagger i} \nonumber\\
&& +\partial_{\mu}\bar{D}\partial^{\mu}\bar{D}^{\dagger}-m^2\bar{D}\bar{D}^{\dagger}-\partial_{\mu}\bar{D}^*_{i}\partial^{\mu}\bar{D}^{*\dagger i}+m^2\bar{D}^*_{i}\bar{D}^{*\dagger i} \nonumber\\
&& -\frac{1}{2}\frac{m\Delta_m}{f_{\pi}}[\bar{D}_0^*(M+M^{\dagger})\bar{D}_0^{*\dagger}-\bar{D}_{1i}(M+M^{\dagger})\bar{D}_1^{\dagger i}-\bar{D}(M+M^{\dagger})\bar{D}^{\dagger}+\bar{D}_{i}^*(M+M^{\dagger})\bar{D}^{*i\dagger}] \nonumber\\
 && -\frac{1}{2}\frac{m\Delta_m}{f_{\pi}}[\bar{D}_0^*(M-M^{\dagger})\bar{D}^{\dagger}-\bar{D}_{1i}(M-M^{\dagger})\bar{D}^{*\dagger i}-\bar{D}(M-M^{\dagger})\bar{D}_0^{*\dagger}+\bar{D}_{i}^{*}(M-M^{\dagger})\bar{D}_1^{\dagger i}]\nonumber\\ 
&& -\frac{g}{2}\frac{m}{f_{\pi}}[\bar{D}_1^{i}(\partial_{i}M^{\dagger}-\partial_{i}M)\bar{D}_0^{*\dagger}-\bar{D}_0^{*}(\partial_{i}M^{\dagger}-\partial_{i}M)\bar{D}_1^{\dagger i}-\frac{1}{m}\epsilon^{ijk0}\bar{D}_{1i}(\partial_{j}M^{\dagger}-\partial_{j}M)i\partial_{0}\bar{D}_{1k}^{\dagger}] \nonumber\\
&& +\frac{g}{2}\frac{m}{f_{\pi}}[\bar{D}^{*i}(\partial_{i}M^{\dagger}-\partial_{i}M)\bar{D}^{\dagger}-\bar{D}(\partial_{i}M^{\dagger}-\partial_{i}M)\bar{D}^{*\dagger i}-\frac{1}{m}\epsilon^{ijk0}\bar{D}_{i}^*(\partial_{j}M^{\dagger}-\partial_{\nu}M)i\partial_{0}\bar{D}_{k}^{*\dagger}] \nonumber\\
&& +\frac{g}{2}\frac{m}{f_{\pi}}[\bar{D}_1^{i}(\partial_{i}M^{\dagger}+\partial_{i}M)\bar{D}^{\dagger}+\bar{D}(\partial_{i}M^{\dagger}+\partial_{i}M)\bar{D}_1^{\dagger i}]\nonumber\\
& & -\frac{g}{2}\frac{m}{f_{\pi}}[\bar{D}_0^*(\partial_{i}M^{\dagger}+\partial_{i}M)\bar{D}^{*\dagger i}+\bar{D}^{*i}(\partial_{i}M^{\dagger}+\partial_{i}M)\bar{D}_0^{*\dagger}]\nonumber\\
& & -\frac{g}{2}\frac{1}{f_{\pi}}[\epsilon^{ijk0}\bar{D}_{1j}(\partial_{k}M^{\dagger}+\partial_{k}M)i\partial_{0}\bar{D}_{i}^{*\dagger}+\epsilon^{ijk0}\bar{D}_{i}^*(\partial_{k}M^{\dagger}+\partial_{k}M)i\partial_{0}\bar{D}_{1j}^{\dagger}] \ , \label{Lagrangian}
\end{eqnarray}
\end{widetext}
where we used $\partial_0^2+m^2\sim O(\Lambda_{\rm QCD}m)$ ($\partial_0$ acts on ``$\bar{D}$ mesons''), set the time derivative of the pion field $U$ to be zero, and neglected
the corrections of $O(\Lambda_{\rm QCD}^2)$.


\section{Equations of Motion for ``$\bar{D}$ mesons'' in the DCDW}
\label{sec:EOMs}

In this section, we derive the EoMs for ``$\bar{D}$ mesons'' in the 
DCDW
from the Lagrangian~(\ref{Lagrangian}). 
In the present analysis, we use the following form of the classical configuration for the  
chiral field $M$ in the DCDW
(See, e.g. \cite{Buballa:2014tba}): 
\begin{eqnarray}
M = \phi\, {\rm cos}(2fx) + i\tau^3\phi\, {\rm sin}(2fx) \ ,\label{CDWPion}
\end{eqnarray}
where $x$ is taken to be the direction of the density wave. $f$ is the wave number of this modulation and $\phi$ is the magnitude of the chiral condensate. 
From the Lagrangian~(\ref{Lagrangian}) with this configuration inserted, we get EoMs of ``$\bar{D}$ mesons'' in the DCDW as
\begin{widetext}
\begin{eqnarray}
&& (\partial_{\mu}\partial^{\mu}+m^2)\bar{D}-m\Delta_m\tilde{\phi}\, {\rm cos}(2fx)\bar{D}+im\Delta_m\tilde{\phi}\, {\rm sin}(2fx)\bar{D}_0^*\tau^3+2ifgm\tilde{\phi}\, {\rm cos}(2fx) \bar{D}^{*1}\tau^3 \nonumber\\
&& +2fgm\tilde{\phi}\, {\rm sin}(2fx)\bar{D}_{1}^1 = 0 \ , \nonumber\\\nonumber\\
&& (\partial_{\mu}\partial^{\mu}+m^2)\bar{D}^{* i}-m\Delta_m\tilde{\phi}\, {\rm cos}(2fx)\bar{D}^{*i}+im\Delta_m\tilde{\phi}\, {\rm sin}(2fx)\bar{D}_{1}^i\tau^3-2ifgm\tilde{\phi}\, {\rm cos}(2fx)\delta_{i1}\bar{D}\tau^3 \nonumber\\
&& -2ifg\tilde{\phi}\, {\rm cos}(2fx)\epsilon^{1ij}i\partial_0\bar{D}^{*j}\tau^3-2fgm\tilde{\phi}\, \delta_{i1}{\rm sin}(2fx)\bar{D}_{0}^*-2fg\tilde{\phi}\, {\rm sin}(2fx)\epsilon^{1ij}i\partial_0\bar{D}_{1}^j = 0 \ , \nonumber\\\nonumber\\
&&  (\partial_{\mu}\partial^{\mu}+m^2)\bar{D}_{0}^*+m\Delta_m\tilde{\phi}\, {\rm cos}(2fx)\bar{D}_{0}^*-im\Delta_m\tilde{\phi}\, {\rm sin}(2fx)\bar{D}\tau^3-2ifgm\tilde{\phi}\, {\rm cos}(2fx)\bar{D}_{1}^{1}\tau^3  \nonumber\\
&& -2fgm\tilde{\phi}\, {\rm sin}(2fx)\bar{D}_{}^{*1}= 0 \ ,\nonumber\\\nonumber\\
&& (\partial_{\mu}\partial^{\mu}+m^2)\bar{D}_{1}^{i}+m\Delta_m\tilde{\phi}\, {\rm cos}(2fx)\bar{D}_{1}^{i}-im\Delta_m\tilde{\phi}\, {\rm sin}(2fx)\bar{D}_{}^{*i}\tau^3+2ifgm\tilde{\phi}\, {\rm cos}(2fx)\delta_{i1}\bar{D}_{0}^*\tau^3 \nonumber\\
&& +2ifg\tilde{\phi}\, {\rm cos}(2fx)\epsilon^{1ij}i\partial_0\bar{D}_{1}^j\tau^3+2fgm\tilde{\phi}\, {\rm sin}(2fx)\delta_{i1}\bar{D}+2fg\tilde{\phi}\, {\rm sin}(2fx)\epsilon^{1ij}i\partial_0\bar{D}^{*j}= 0\ , \label{EoMs}
\end{eqnarray}
\end{widetext}
where we defined $\tilde{\phi}=\phi/f_{\pi}$.
Note that, charged pions do not condense, then 
``$\bar{D}$ mesons'' with different charges do not mix with each other.
In the following, we consider only the ``neutral $\bar{D}$ mesons'' composed of an anti-charm quark and an up quark, denoted as $\bar{D}_u$ ($J^P=0^-$), $\bar{D}_u^{\ast i}$ ($J^P=1^-$), $\bar{D}_{0u}^{\ast}$ ($J^P=0^+$) and $\bar{D}_{1u}^i$ ($J^P=1^+$) with $i=1,2,3$.
Furthermore, these eight states are separated into two sectors:
The mesons in ($\bar{D}_u$, $\bar{D}^{\ast 1}_u$, $\bar{D}_{0\,u}^\ast$, $\bar{D}_{1\,u}^{1}$) are mixed among themselves, while they do not mix with any of ($\bar{D}_u^{\ast 2}$, $\bar{D}^{\ast 3}_u$, $\bar{D}_{1\,u}^{2}$, $\bar{D}_{1\,u}^{3}$). This situation is the same for the charged ``$\bar{D}$ mesons'' composed of an anti-charm quark and a down quark, i.e., ($\bar{D}_d,\bar{D}_d^{*i}, \bar{D}_{0\,d}^*,\bar{D}_{1\,d}^i)$. 
We would like to note that dispersion relations for the ``charged $\bar{D}$ mesons'' are exactly the same as those for the ``neutral $\bar{D}$ mesons'' because of the $Z_2$ symmetry pointed in~\cite{Suenaga:2014dia}:
In the DCDW, the condensate is invariant
under the $\mbox{U}(1)_l \times \mbox{U}(1)_I \times Z_2$ symmetry transformation,
where $\mbox{U}(1)_l $ is a subgroup of the light-spin $\mbox{SU}(2)_l$ symmetry and $\mbox{U}(1)_I$ is of the isospin $\mbox{SU}(2)_I$ symmetry. $Z_2$ is a group associated with the ``charge conjugation''
under which both $\mbox{U}(1)_l $ and $\mbox{U}(1)_I$ charges are inverted
simultaneously.

In order to get the dispersion relations 
in the present system, we need to obtain the EoMs of the ``neutral $\bar{D}$ mesons'' in the momentum space. 
Since the potential is periodic with the period of $\pi/f$ in the DCDW, 
we employ the Bloch's theorem. 
The Bloch's theorem is shortly summarized in Appendix~\ref{sec:BlochTheorem}. 

\begin{widetext}
The Fourier expansions of the ``neutral $\bar{D}$ mesons'' are expressed as
\begin{eqnarray}
\bar{D}_{\rm neutral}^s(x) = \sum_{k_x,k_y,k_z}\sum_{K'}C_{k_x-K'}^s{\rm e}^{i(k_x-K')x}{\rm e}^{-iEt+ik_yy+ik_zz}\ ,
\label{Fourier}
\end{eqnarray}
where the superscript $s$ runs as $s =1,2$, and $\bar{D}_{\rm neutral}^s$ and $C_{k_x-K'}^s$ are 
column vectors with four components for the  ``neutral $\bar{D}$ mesons'' fields and their Fourier coefficients. 
They for $s=1$ are expressed
as
\begin{eqnarray}
\bar{D}_{\rm neutral}^{s=1}(x) = \left(
\begin{array}{c}
\bar{D}_u(x) \\
\bar{D}^{\ast 1}_u(x)\\
\bar{D}_{0\,u}^\ast(x)\\
\bar{D}_{1\,u}^{1}(x) \\
\end{array}
\right)\ ,\ \ 
C_{k_x-K'}^{s=1} = \left(
\begin{array}{c}
C_{k_x-K'}^{\bar{D}_u} \\
C_{k_x-K'}^{\bar{D}^{\ast 1}_u}\\
C_{k_x-K'}^{\bar{D}_{0\,u}^\ast}\\
C_{k_x-K'}^{\bar{D}_{1\,u}^{1}} \\
\end{array}
\right)\ , 
\end{eqnarray}
and 
for $s=2$ as
\begin{eqnarray}
\bar{D}_{\rm neutral}^{s=2}(x) = \left(
\begin{array}{c}
\bar{D}_u^{\ast 2}(x) \\
\bar{D}^{\ast 3}_u(x)\\
\bar{D}_{1\,u}^2(x)\\
\bar{D}_{1\,u}^{3}(x) \\
\end{array}
\right)\ ,\ \ 
C_{k_x-K'}^{s=2} = \left(
\begin{array}{c}
C_{k_x-K'}^{\bar{D}_u^{\ast 2}} \\
C_{k_x-K'}^{\bar{D}^{\ast 3}_u}\\
C_{k_x-K'}^{\bar{D}_{1\,u}^2}\\
C_{k_x-K'}^{\bar{D}_{1\,u}^{3}} \\
\end{array}
\right)\ . 
\end{eqnarray}
\end{widetext}
In Eq.~(\ref{Fourier}), 
$k_x$ denotes the crystal momentum along the $x$ axis which is taken to be the direction of the density wave.
This $k_x$ lies only in the first Brillouin zone and $K'$ is the reciprocal lattice vector which is integral multiple of $K=2f$.  While $k_y$ and $k_z$ are the momenta along the $y$ and $z$ axis, respectively, for which there are no Brillouin zones.
$C^s_{k_x-K'}$ is a
 column vector for fixed momenta ($k_x - K'$, $k_y$, $k_z$), where we omit the indices of $k_y$ and $k_z$ for notational simplicity.
The EoMs of the ``neutral $\bar{D}$ mesons'' in the momentum space are written as the matrix form of
\begin{eqnarray}
\left(
\begin{array}{ccccccc}
\ddots & \ddots & 0 & 0 & 0   \\
\ddots & \Delta_{k_x-K} & V^s & 0 & 0 \\
0 & (V^s)^{\dagger}  & \Delta_{k_x} & V^s & 0 \\
0 & 0 & (V^s)^{\dagger}& \Delta_{k_x+K} & \ddots  \\ 
0 & 0 & 0 & \ddots & \ddots 
\\
\end{array}
\right) \left(
\begin{array}{c}
\vdots \\
C^s_{k_x-K} \\
C^s_{k_x}\\
C^s_{k_x+K} \\
\vdots \\
\end{array}
\right) = 0\ , \nonumber\\
\label{EoMsMatrix}
\end{eqnarray}
where
\begin{eqnarray}
\Delta_{q} &=& \left(
\begin{array}{cccc}
-\bar{E}^2+q^2 & 0 & 0 & 0 \\
0 & -\bar{E}^2+q^2 & 0 & 0\\
0 & 0 & -\bar{E}^2+q^2 & 0 \\
0 & 0 & 0 & -\bar{E}^2+q^2 \\
\end{array}
\right)\ , \nonumber\\
\end{eqnarray}
and $\bar{E}^2$ is defined as $\bar{E}^2 = E^2-k_{\perp}^2-m^2$ ($k_{\perp}^2=k_y^2+k_z^2$). 
The matrix $V^s$ is given as
\begin{eqnarray}
V^{s=1} &=& \left(
\begin{array}{cccc}
-\frac{m\Delta_m}{2}\tilde{\phi} & ifgm\tilde{\phi} & -\frac{m\Delta_m}{2}\tilde{\phi} & ifgm\tilde{\phi} \\
-ifgm\tilde{\phi} & -\frac{m\Delta_m}{2}\tilde{\phi} & -ifgm\tilde{\phi} & -\frac{m\Delta_m}{2}\tilde{\phi} \\
\frac{m\Delta_m}{2}\tilde{\phi} & -ifgm\tilde{\phi} & \frac{m\Delta_m}{2}\tilde{\phi} & -ifgm\tilde{\phi} \\
ifgm\tilde{\phi} & \frac{m\Delta_m}{2}\tilde{\phi} & ifgm\tilde{\phi} & \frac{m\Delta_m}{2}\tilde{\phi} \\ \label{V1}
\end{array}
\right) 
\nonumber\\
\label{Matrix1}
\end{eqnarray}
for $s=1$, 
and 
\begin{eqnarray}
V^{s=2} &=& \left(
\begin{array}{cccc}
-\frac{m\Delta_m}{2}\tilde{\phi} & ifgE\tilde{\phi} & -\frac{m\Delta_m}{2}\tilde{\phi} & ifgE\tilde{\phi} \\
-ifgE\tilde{\phi} & -\frac{m\Delta_m}{2}\tilde{\phi} & -ifgE\tilde{\phi} & -\frac{m\Delta_m}{2}\tilde{\phi} \\
\frac{m\Delta_m}{2}\tilde{\phi} & -ifgE\tilde{\phi} & \frac{m\Delta_m}{2}\tilde{\phi} & -ifgE\tilde{\phi} \\
ifgE\tilde{\phi} & \frac{m\Delta_m}{2}\tilde{\phi} & ifgE\tilde{\phi} & \frac{m\Delta_m}{2}\tilde{\phi} \\
\end{array}
\right)
\nonumber\\\label{Matrix2}
\end{eqnarray}
for $s=2$. We would like to stress that the present form of the potentials in the position space is expressed by only $\cos (2fx)$ and $\sin(2fx)$ provided by~(\ref{CDWPion}), so that the off-diagonal components of the matrix in the left-hand-side of Eq.~(\ref{EoMsMatrix}) connects only the nearest Brillouin zones.

Note that, when we take the heavy quark mass to infinity, $E/m \to 1$, and hence the EoMs for ($\bar{D}_u^{\ast 2}$, $\bar{D}^{\ast 3}_u$, $\bar{D}_{1\,u}^{2}$, $\bar{D}_{1\,u}^{3}$) agree with those for ($\bar{D}_u$, $\bar{D}^{\ast 1}_u$, $\bar{D}_{0\,u}^\ast$, $\bar{D}_{1\,u}^{1}$).
This is a consequence of the heavy quark symmetry.
Then, we can easily obtain the dispersion relations of ``$\bar{D}$ mesons'' by diagonalizing the above EoMs.
\\

\section{Dispersion Relations 
in the DCDW}
\label{sec:Dispersions}

In this section, we study dispersion relations for ``$\bar{D}$ mesons'' 
in the DCDW phase. 
As we stated in the previous section, in the present case, ``$\bar{D}$ mesons" including an up quark do not mix with the ones with a down quark.  
Furthermore, the mesons in ($\bar{D}_u$, $\bar{D}^{\ast 1}_u$, $\bar{D}_{0\,u}^\ast$, $\bar{D}_{1\,u}^{1}$) are mixed among themselves, separately from the ones of ($\bar{D}_u^{\ast 2}$, $\bar{D}^{\ast 3}_u$, $\bar{D}_{1\,u}^{2}$, $\bar{D}_{1\,u}^{3}$).
Since the EoMs for ($\bar{D}_u^{\ast 2}$, $\bar{D}^{\ast 3}_u$, $\bar{D}_{1\,u}^{2}$, $\bar{D}_{1\,u}^{3}$) agree with those for ($\bar{D}_u$, $\bar{D}^{\ast 1}_u$, $\bar{D}_{0\,u}^\ast$, $\bar{D}_{1\,u}^{1}$) in the the heavy quark limit,
we consider only 
($\bar{D}_u$, $\bar{D}_u^{*1}$, $\bar{D}_{0\, u}^*$, $\bar{D}_{1\, u}^1$) in this section.
In the following, we 
first compute the EoMs with $g=0$ to show 
the features of the medium modified modes,  
and to study the dependence of our results on 
the magnitude of the chiral condensate. 
Next, we study the mixing effects among these four states with $g\neq0$. Finally, we include the effect of mass difference 
between $\bar{D}$ and $\bar{D}^\ast$ mesons as well as that between $\bar{D}_0^\ast$ and $\bar{D}_1$ mesons 
which breaks 
the heavy quark symmetry.

\subsection{Dispersion Relations with $g=0$ }
\label{sec:MHMGWithoutG}

We study dispersion relations for 
($\bar{D}_u$, $\bar{D}_u^{*1}$, $\bar{D}_{\, u}^*$, $\bar{D}_{1\, u}^1$) mesons with setting $g=0$. In this case, we notice that  these four states are separated into two sectors  
as ($\bar{D}_u$, $\bar{D}_{0\, u}^{*}$) and  
($\bar{D}_u^{*1}$, $\bar{D}_{1\, u}^1$),
and the EoMs for these two sectors are identical. This means that the system has the $Z_2$ symmetry for flipping the spin of Brown Muck~\footnote{Spin of the Brown Muck is $1/2$ in the present analysis. }. 
Then we obtain doubly degenerated dispersion relations here. In the following we consider only the dispersion relations in the ($\bar{D}_u$, $\bar{D}_{0\, u}^{*}$) sector.

Here we take $g=0$, so that the two-by-two matrix for the ($\bar{D}_u$, $\bar{D}_{0\, u}^{*}$) sector of $V^{s=1}$ in Eq.~(\ref{V1}) is reduced to 
\begin{equation}
V = \begin{pmatrix}
- v & - v \\ v & v \\
\end{pmatrix}
\ ,
\end{equation}
where $v= \frac{m \Delta m}{2} \tilde{\phi}$.
It should be noticed that the form of this matrix reflects
the chiral partner structure, in addition to the fact that the potential in the position space is written in terms of $\cos(2fx)$ and $\sin(2fx)$.
Then, 
as is explained in detail in Appendix~\ref{sec:ExtendedZone},
we can diagonalize the matrix in Eq.~(\ref{EoMsMatrix}) to determine the infinite numbers of the dispersion relations as
\\
\begin{widetext}
\begin{eqnarray}
E^2 &=& m^2+\frac{1}{2}\left[(k_x+(n+1)K)^2+(k_x+nK)^2\right] + k_{\perp}^2 
\pm\frac{1}{2}\sqrt{\left[(k_x+(n+1)K)^2-(k_x+nK)^2\right]^2+4\tilde{\phi}^2m^2\Delta_m^2}\ , \nonumber\\\label{Eigenvalues}
\end{eqnarray}
where $n$ is an integer. The corresponding eigenvectors, up to the normalization factors, are expressed as
\begin{equation}
\begin{pmatrix}
\vdots \\ 0 \\ 0\\ C_{k_x + n K} \\ C_{k_x + (n+1) K} \\ 0 \\ 0\\ \vdots \\ 
\end{pmatrix}
\ , \label{EigenvectorMS}
\end{equation}
where
\begin{align}
C_{k_x + n K} & = \begin{pmatrix} C_{k_x + n K}^{\bar{D}_u} \\ C_{k_x + n K}^{\bar{D}_{0\, u}^\ast} \\ \end{pmatrix} 
=  \begin{pmatrix} \displaystyle
\frac{ 4 v} { \left(k_x + n K\right)^2 - \left(k_x + (n+1)K\right)^2 \mp \sqrt{ \left\{ \left(k_x + n K\right)^2 - \left(k_x + (n+1)K\right)^2 \right\}^2 + 16 v^2 } } \\
\displaystyle
-\frac{  4 v} { \left(k_x + n K\right)^2 - \left(k_x + (n+1)K\right)^2 \mp \sqrt{ \left\{ \left(k_x + n K\right)^2 - \left(k_x + (n+1)K\right)^2 \right\}^2 + 16 v^2 } } \\
\end{pmatrix}
\ , \notag\\
C_{k_x + (n+1)K} & = \begin{pmatrix} C_{k_x + (n+1) K}^{\bar{D}_u} \\ C_{k_x + (n+1) K}^{\bar{D}_{0\, u}^\ast} \\ \end{pmatrix}
= \begin{pmatrix} 1 \\ 1 \\ \end{pmatrix} 
 \label{Eigenvectors}
\end{align}

\end{widetext}

\begin{figure*}[t]
  \begin{center}
    \begin{tabular}{c}
    
 \subfigure[\ $f=100$ MeV]{
      \begin{minipage}{0.32\hsize}
        \begin{center}
         \includegraphics*[scale=0.5]{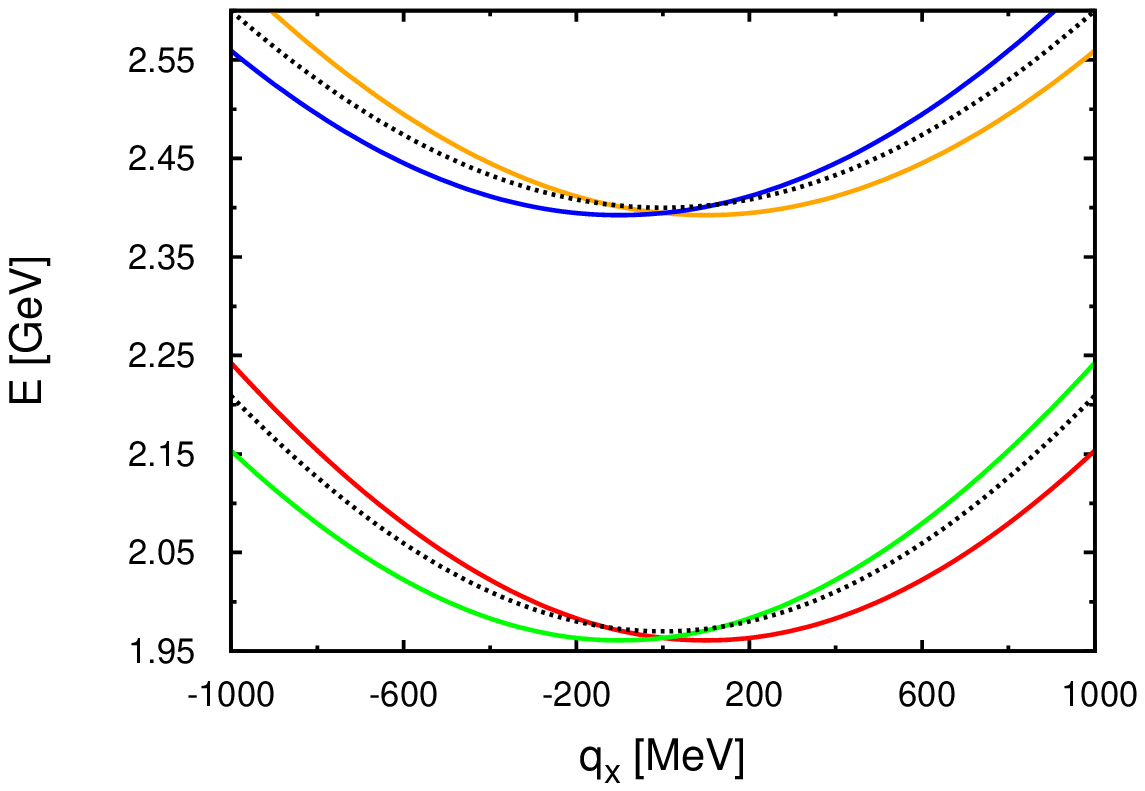}
          \hspace{0.5cm} 
        \end{center}
      \end{minipage}
}
 \subfigure[\ $f=200$ MeV]{
      \begin{minipage}{0.32\hsize}
        \begin{center}
          \includegraphics*[scale=0.5]{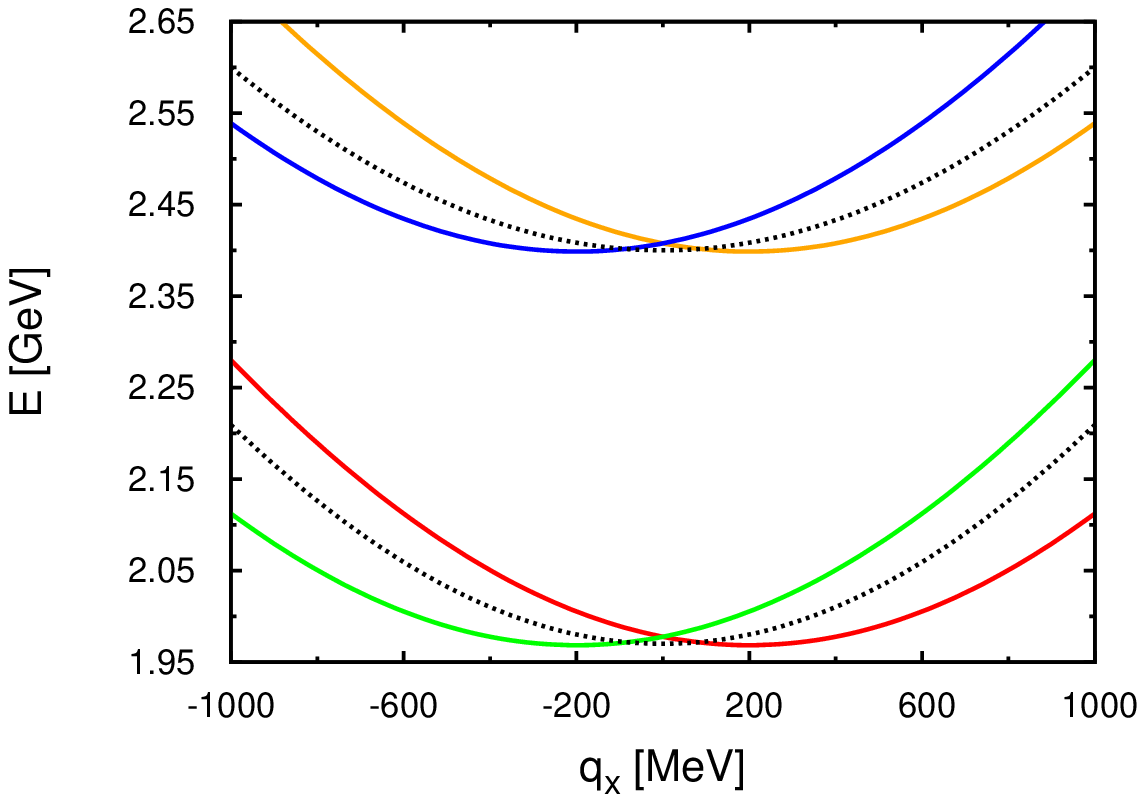}
          \hspace{0.5cm}
        \end{center}
      \end{minipage}
}
 \subfigure[\ $f=400$ MeV]{
      \begin{minipage}{0.32\hsize}
        \begin{center}
          \includegraphics*[scale=0.5]{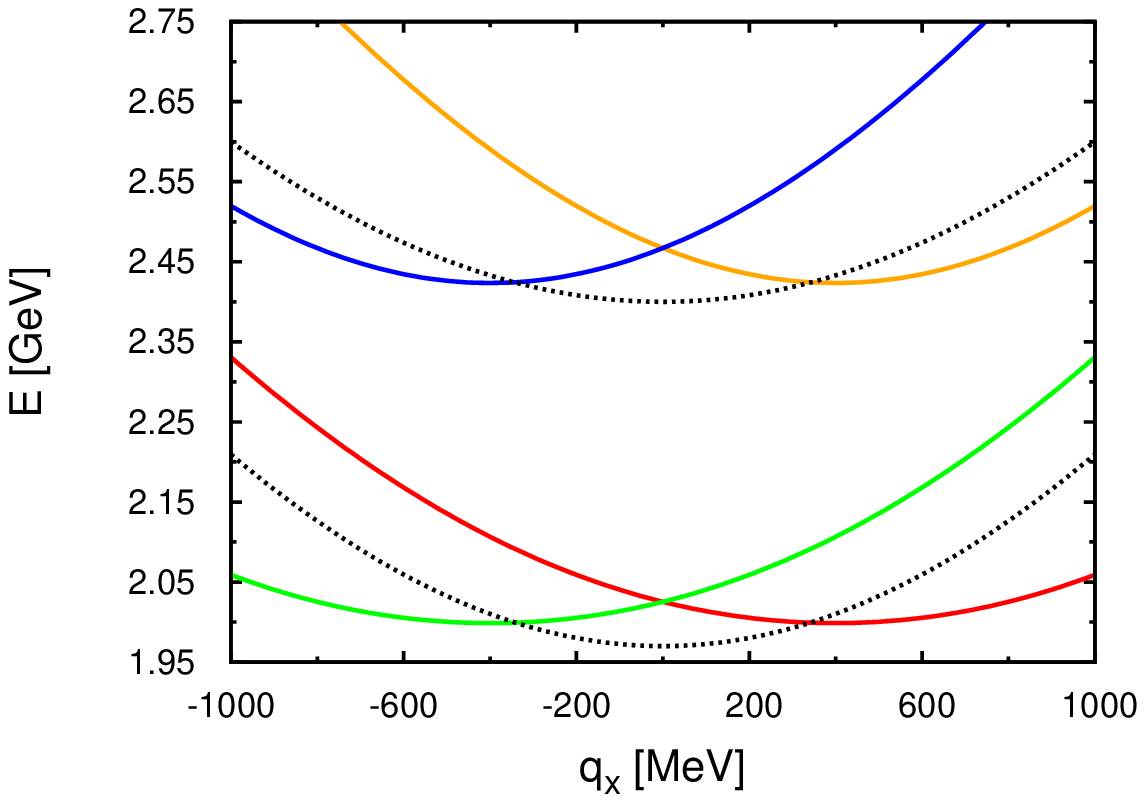}
          \hspace{0.5cm} 
        \end{center}
      \end{minipage}
}
    \end{tabular}
   \caption{(color online) Dispersion relations 
for (a)  $f=100$ MeV, (b) $f=200$ MeV and (c) $f=400$ MeV 
with $k_{\perp}^2=0$ and $\tilde{\phi}=1$. 
Solid curves show the results obtained in the present analysis, while the dotted curves are the dispersion relations for the $\bar{D}$ and $\bar{D}_0^*$ mesons in the vacuum.}
\label{DispersionWithoutG}
  \end{center}
\end{figure*}

(The double-sign in $C_{k_x+nK}$ corresponds to the double-sign in  eigenvalues~(\ref{Eigenvalues})). The above 
dispersion relations are written in terms of the crystal momentum $k_x$.
Here we rewrite $k_x$ in terms of familiar definition of momentum $q_x$, which is defined via the Fourier expansion of the ``neutral $\bar{D}$ mesons'' as
\begin{eqnarray}
\bar{D}_{\rm neutral}^s(x) =\sum_{q_x,k_y,k_z}C_{q_x}^s{\rm e}^{-iEt+iq_xx+ik_yy+ik_zz}\ .
\end{eqnarray}
Using the fact that the eigenvector has non-vanishing values only in $C_{k_x+nK}$ and $C_{k_x+(n+1)K}$ for $n$-th band as in Eq.~(\ref{EigenvectorMS}), we change the variable as $q_x=k_x+(n+1)K$ or $q_x=k_x+nK$ to obtain the four dispersion relations for ($\bar{D}_u$, $\bar{D}_{0\, u}^{*}$) as
\begin{eqnarray}
E^2 &=& m^2+\frac{1}{2}\left[(q_x+K)^2+q_x^2\right] + k_{\perp}^2 \nonumber\\
&& \pm\frac{1}{2}\sqrt{\left[(q_x+K)^2-q_x^2\right]^2+4\tilde{\phi}^2m^2\Delta_m^2}\ , \label{Dispersion11}
\\
E^2 &=& m^2+\frac{1}{2}\left[q_x^2+(q_x-K)^2\right] + k_{\perp}^2 \nonumber\\
&& \pm\frac{1}{2}\sqrt{\left[q_x^2-(q_x-K)^2\right]^2+4\tilde{\phi}^2m^2\Delta_m^2}\ . \label{Dispersion12}
\end{eqnarray}
This procedure is explained in detail in appendix.~\ref{sec:ExtendedZone}. We plot these dispersion relations 
with $k_{\perp}^2=0$ and $\tilde{\phi}=1$ for $f=100$ MeV, $f=200$ MeV and $f=400$ MeV 
by the solid curves
in Fig.~\ref{DispersionWithoutG}.
For comparison we also plot  the dispersion relations for the $\bar{D}$ and $\bar{D}_0^*$ mesons in the vacuum by
the dotted curves. The blue and green solid curves show the `+' sign and `-' sign in the right hand side (RHS) of the dispersion relations in Eq.~(\ref{Dispersion11}), respectively, while the orange and red curves are for the `+' and `-' sign in the RHS of Eq.~(\ref{Dispersion12}), respectively. 

Here we consider the group velocity in the $x$-direction at the low-momentum limit, $v_x = \left. \frac{d E}{d q_x} \right\vert_{q_x = k_\perp^2=0}$. 
The dispersion relations~(\ref{Dispersion11}) and~(\ref{Dispersion12}) lead to  
\begin{align}
v_x &= \frac{K}{2E} \left[ 1 \pm \frac{K^2}{ \sqrt{ K^4 + 4 \tilde{\phi}^2 m^2 \Delta_m^2 } } \right] \ , 
\label{velocity1}\\
v_x &= - \frac{K}{2E} \left[ 1 \pm \frac{K^2}{ \sqrt{ K^4 + 4 \tilde{\phi}^2 m^2 \Delta_m^2 } } \right] 
\ , \label{velocity2}
\end{align}
respectively.
Since $1 \pm \frac{K^2}{ \sqrt{ K^4 + 4 \tilde{\phi}^2 m^2 \Delta_m^2 } }  >0$, 
these implies that the dispersion relations drawn by the green and blue curves have the positive velocity at $q_x=0$ shown by Eq.~(\ref{velocity1}), 
while those by the red and orange curves have the negative velocity 
by Eq.~(\ref{velocity2}). 
Non-vanishing velocity at zero-momentum limit contrasts to the velocity of the free particle, e.g. obtained from the dispersion relations shown by dotted curves.
This implies that all the modes shown by colored curves in Fig.~\ref{DispersionWithoutG} are medium modified modes which do not exist in the vacuum. 
The red and orange curves are minimized at $q_x=f$, while the green and blue curves are minimized at $q_x=-f$. 
We notice that the feature that the velocity is opposite to the momentum and the minimum of energy is realized for non-zero momentum.

When $K$ is zero, 
the dispersion relations with `$+$' sign in Eqs.~(\ref{Dispersion11}) and (\ref{Dispersion12})  agree with those of the free $\bar{D}_{0u}^\ast$ meson,  and the ones  with `$-$' sign agree with those of free $\bar{D}_u$ meson.  
By increasing $K=2f$ from zero, the mode corresponding to $\bar{D}_{0u}^\ast$ meson split into two modes expressed by the blue and orange curves in Fig.~\ref{DispersionWithoutG}, while the mode to $\bar{D}$ meson into those by the green and red curves, 
due to the mixing between $\bar{D}_u$ and $\bar{D}_{0\, u}^*$ mesons and periodicity of 
the potential.
When we take the short wavelength limit, $K^2 \gg q_x^2,\, 2\tilde{\phi}m\Delta_m$, 
the energies
corresponding to the blue and orange curves 
are approximated as
$E^2 \sim m^2 + K^2 + 2 K q_x + q_x^2 + k_\perp^2$
and 
$E^2 \sim m^2 + K^2 - 2 K q_x + q_x^2 + k_\perp^2$.
On the other hand,
those 
 to the green and red curves 
are approximated as
$E^2 = m^2 + q_x^2 + k_\perp^2$.
The latter 
two modes follow the dispersion relations of the free particle with the degenerate mass of 
$m$, which correspond to the ones obtained in Ref.~\cite{Suenaga:2014sga} by taking the space average of the inhomogeneous condensate in the half-Skyrmion phase.

We also note that energy gaps do not appear although we employ the Bloch's theorem, which presumably reflects the chiral partner structure and the heavy quark symmetry.

\begin{figure*}[t]
  \begin{center}
    \begin{tabular}{c}

 \subfigure[\ $\tilde{\phi}=0.6$]{
      \begin{minipage}{0.32\hsize}
        \begin{center}
         \includegraphics*[scale=0.5]{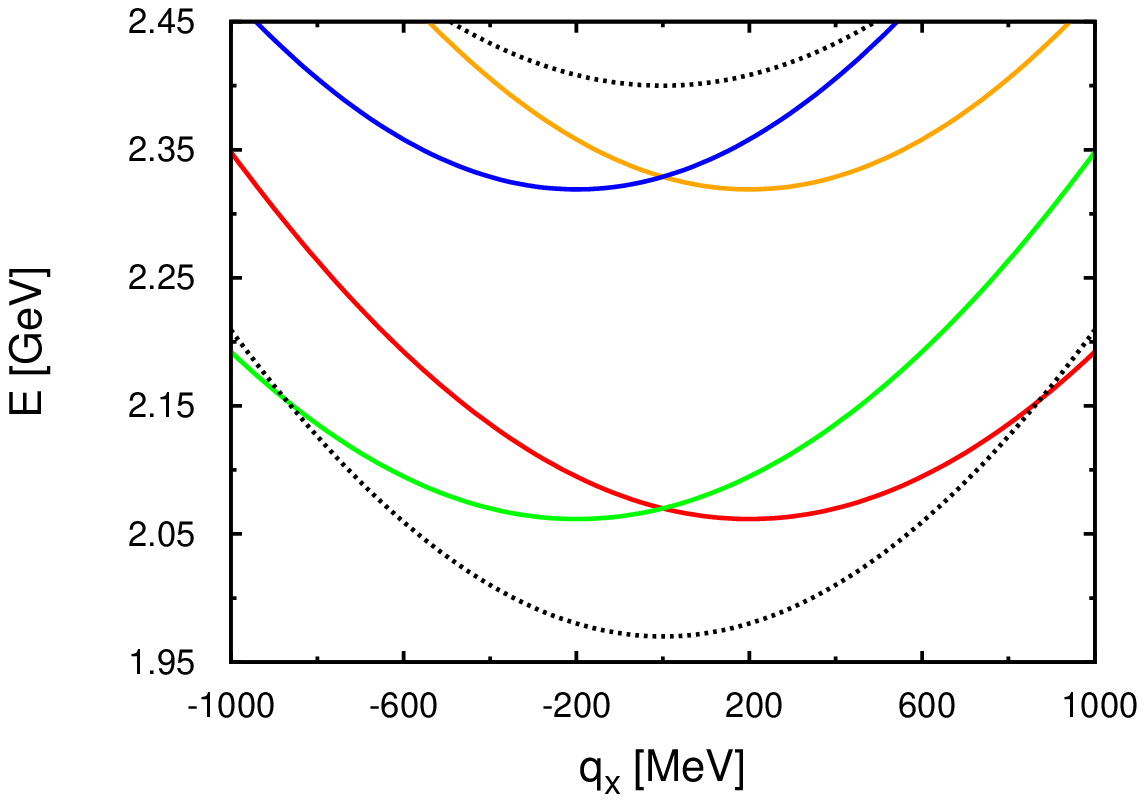}
          \hspace{0.5cm} 
        \end{center}
      \end{minipage}
}
  \subfigure[\ $\tilde{\phi}=0.3$]{
      \begin{minipage}{0.32\hsize}
        \begin{center}
          \includegraphics*[scale=0.5]{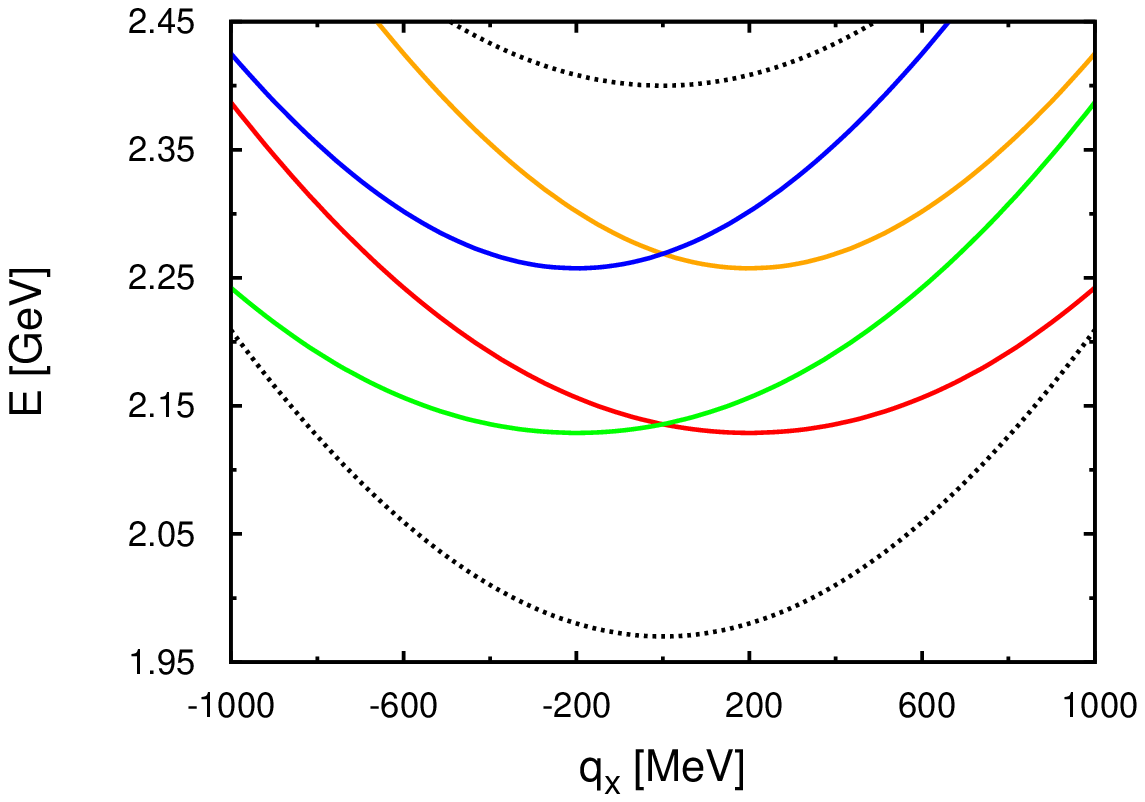}
          \hspace{0.5cm}
        \end{center}
      \end{minipage}
}
 \subfigure[\ $\tilde{\phi}=0$]{
      \begin{minipage}{0.32\hsize}
        \begin{center}
          \includegraphics*[scale=0.5]{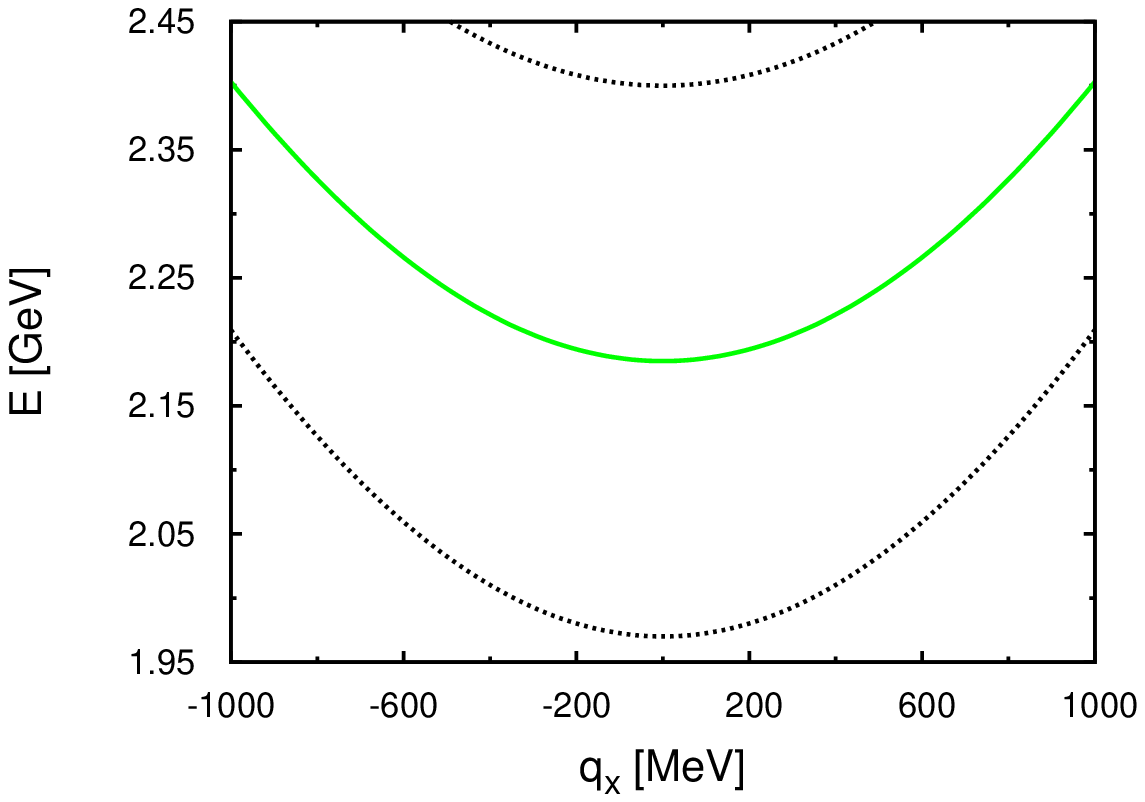}
          \hspace{0.5cm}
        \end{center}
      \end{minipage}
}
    \end{tabular}
  \caption{(color online) Dispersion relations with $f=200$ MeV and $k_{\perp}^2=0$
for (a) $\tilde{\phi}=0.6$\,MeV, (b) $\tilde{\phi}=0.3$\,MeV and (c) $\tilde{\phi}=0$. 
Solid curves show the results obtained in the present analysis, while the dotted curves are for the dispersion relations for the  $\bar{D}$ and $\bar{D}_0^*$ mesons in the vacuum.
In the case of $\tilde{\phi}=0$ in (c), the collective modes disappear and there is a doubly degenerated dispersion relation shown by the green line.}
\label{DispersionPhiDep}
  \end{center}
\end{figure*}

Next, we shall study the dependence of the dispersion relations for ($\bar{D}_u$, $\bar{D}_{0\, u}^*$) on the 
value of $\tilde{\phi}$ which is the ratio of the chiral condensate in the DCDW to that in the vacuum. 
We plot the dispersion relations for $\tilde{\phi}=0.6$, $\tilde{\phi}=0.3$ and $\tilde{\phi}=0$ with $k_{\perp}^2=0$ and $f=200$ MeV in Fig.~\ref{DispersionPhiDep}. 
As the value of 
$\tilde{\phi}$ becomes small, the two modes shown by the red and green curves move up while the two modes by the orange and blue curves move down.
This implies that we can extract the magnitude of 
$\tilde{\phi}$
by measuring 
the difference of the masses of two modes whchi are defined as the energies at $q_x=0$ and $k_\perp^2=0$.
For $\tilde{\phi}=0$, there are only two degenerate modes shown by the green curve in Fig.~\ref{DispersionPhiDep} (c).
The eigenvalues in Eqs.~(\ref{Dispersion11}) and (\ref{Dispersion12}) are reduced to $E^2=m^2+q_x^2+k_\perp^2$, $E^2=m^2+(q_x+K)^2+k_\perp^2$ and $E^2=m^2+q_x^2+k_\perp^2$, $E^2=m^2+(q_x-K)^2+k_\perp^2$, respectively at $\tilde{\phi}=0$. 
On the other hand, all the Fourier coefficients $C_{q_x}^s$ corresponding to  $E^2=m^2+(q_x+K)^2+k_\perp^2$ and $E^2=m^2+(q_x-K)^2+k_\perp^2$ vanish, which implies that these states do not exist as eigenstates.
As a result there are only two modes which follow the same dispersion relation of $E=\sqrt{m^2+q_x^2+k_\perp^2}$ reflecting 
the restoration of the chiral symmetry.

\subsection{Dispersion Relations with $g\neq0$}
\label{sec:MHMG}

Next, we consider ($\bar{D}_u$, $\bar{D}_u^{*1}$, $\bar{D}_{\, u}^*$, $\bar{D}_{1\, u}^1$) states with $g\neq0$.  All four states can mix in this case. By diagonalizing the eigenvalue matrix in~(\ref{EoMsMatrix}) and changing the variable $k_x$ to $q_x$, we obtain eight eigenvalues as
\begin{eqnarray}
E^2 &=& m^2+\frac{1}{2}\left[(q_x+ K)^2+q_x^2\right] + k_{\perp}^2 \nonumber\\
&& -\frac{1}{2}\sqrt{\left[(q_x+ K)^2-q_x^2\right]^2+4\tilde{\phi}^2m^2(\Delta_m\pm2fg)^2}\ ,  \nonumber\\\label{Dispersion21}
\\
E^2 &=& m^2+\frac{1}{2}\left[q_x^2+(q_x-K)^2\right] + k_{\perp}^2 \nonumber\\
&& -\frac{1}{2}\sqrt{\left[q_x^2-(q_x-K)^2\right]^2+4\tilde{\phi}^2m^2(\Delta_m\pm2fg)^2}\ , \nonumber\\
\label{Dispersion22} 
\\
E^2 &=& m^2+\frac{1}{2}\left[(q_x+ K)^2+q_x^2\right] + k_{\perp}^2 \nonumber\\
&& +\frac{1}{2}\sqrt{\left[(q_x+ K)^2-q_x^2\right]^2+4\tilde{\phi}^2m^2(\Delta_m\pm2fg)^2}\ ,  \nonumber\\\label{Dispersion23}
\\
E^2 &=& m^2+\frac{1}{2}\left[q_x^2+(q_x-K)^2\right] + k_{\perp}^2 \nonumber\\
&& +\frac{1}{2}\sqrt{\left[q_x^2-(q_x-K)^2\right]^2+4\tilde{\phi}^2m^2(\Delta_m\pm2fg)^2}\ . \nonumber\\
\label{Dispersion24}
\end{eqnarray}
From this one can take $g>0$ without loss of generality, so that we take $g >0$ for the definiteness of our discussion below.

\begin{figure*}[t]
  \begin{center}
    \begin{tabular}{c}

  \subfigure[\ $f=100$ MeV]{
      \begin{minipage}{0.32\hsize}
        \begin{center}
         \includegraphics*[scale=0.5]{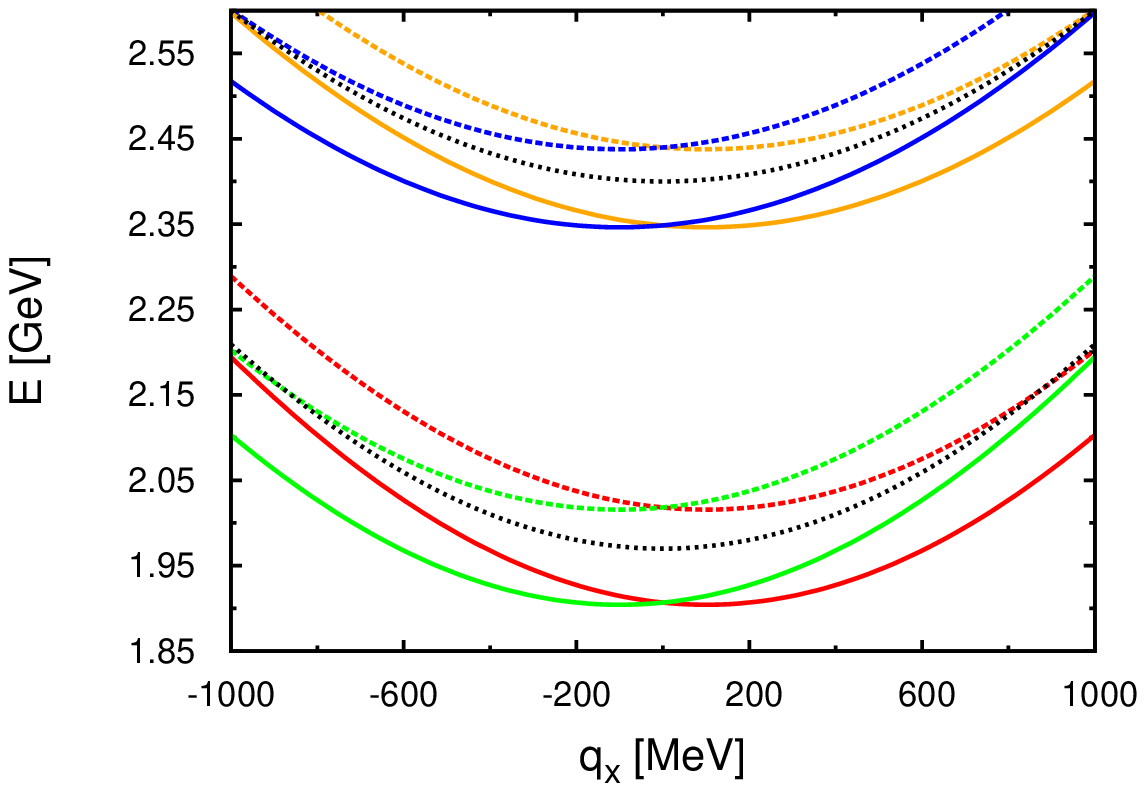}
          \hspace{0.5cm}
        \end{center}
      \end{minipage}
}
     \subfigure[\ $f=200$ MeV]{
      \begin{minipage}{0.32\hsize}
        \begin{center}
          \includegraphics*[scale=0.5]{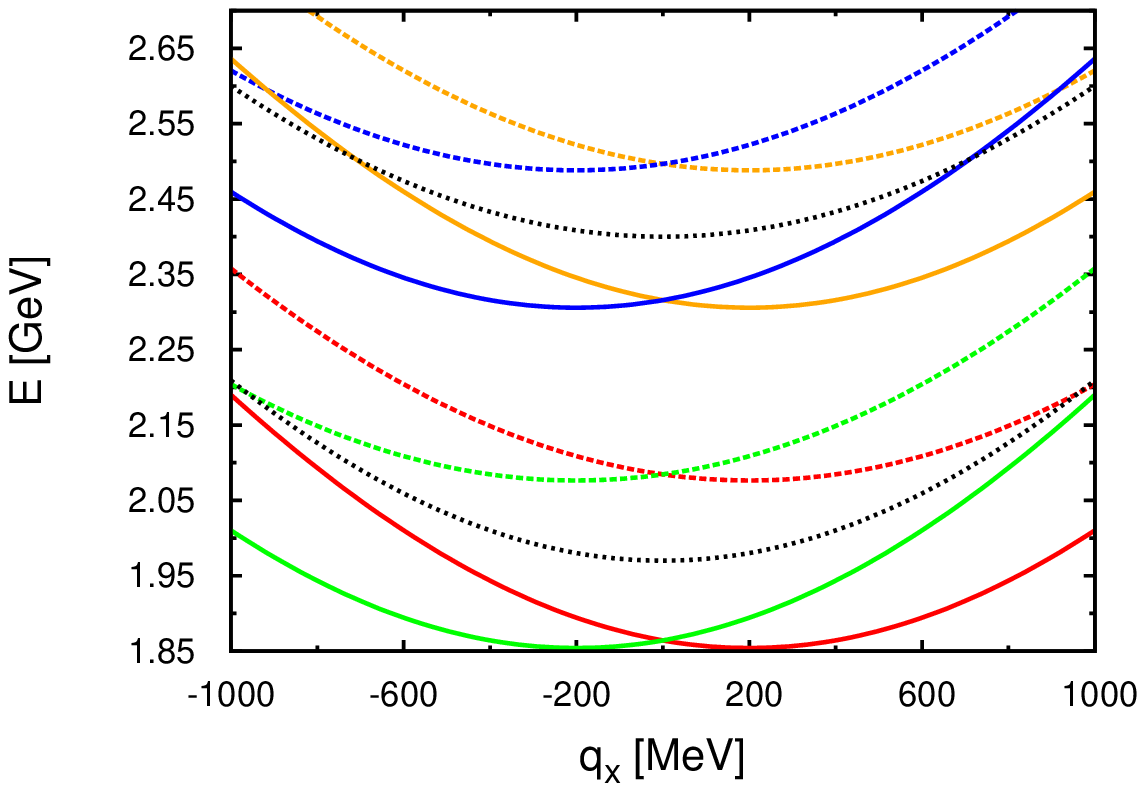}
          \hspace{0.5cm}
        \end{center}
      \end{minipage}
}
 \subfigure[\ $f=400$ MeV]{
      \begin{minipage}{0.32\hsize}
        \begin{center}
          \includegraphics*[scale=0.5]{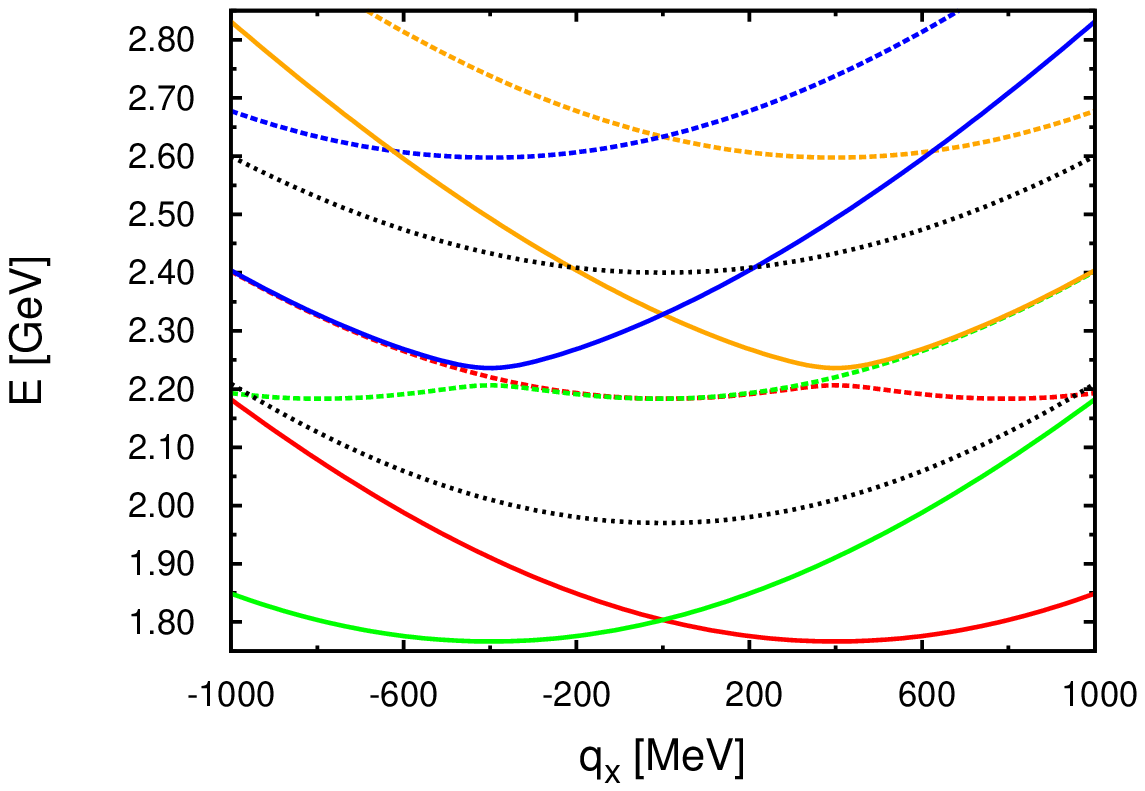}
          \hspace{0.5cm} 
        \end{center}
      \end{minipage}
}
    \end{tabular}
 \caption{(color online) Dispersion relations
for (a) $f=100$ MeV, (b) $f=200$ MeV and (c) $f=400$ MeV with $k_{\perp}^2=0$ and $g=0.50$. 
The green solid curve shows the medium modified dispersion relation given by the eigenvalue with $+2fg$ in Eq.~(\ref{Dispersion21}), while the green dotted curve shows the one with $-2fg$.
The red solid and dotted curves the ones with $+2fg$ and $-2fg$ in Eq.~(\ref{Dispersion22}), 
the blue solid and dotted curves the ones with $-2fg$ and $+2fg$ in Eq.~(\ref{Dispersion23}), 
the orange solid and dotted curves the ones with $-2fg$ and $+2fg$ in Eq.~(\ref{Dispersion24}), respectively.
For comparison, we also plot the dispersion relations for the 
 free $\bar{D}$ and $\bar{D}_0^*$ mesons by black dotted lines.}
\label{Dispersion}
  \end{center}
\end{figure*}

We plot the dispersion relations for $f=100$ MeV, $f=200$ MeV and $f=400$ MeV with $k_{\perp}^2=0$ and $g=0.50$ in Fig.~\ref{Dispersion}. 
We note that the splitting between a solid curve and a dotted curve with the same color is caused by the mixing among spin-0 and spin-1 states pointed in Ref.~\cite{Suenaga:2014dia}. 
In this figure, the green and blue curves have positive slope at $q_x=0$ corresponding to positive velocity,
and the red and orange  curves have negative slope to negative velocity.
All four modes shown by red and orange curves for $f=100$ MeV and $f=200$ MeV take the minimum  energy at $q_x=f$. 
For $f=400$ MeV, on the other hand, the energies for the solid red, solid orange and dotted orange curves are minimized at $q_x=f$, while the one for the dotted red curve is maximized at $q_x=f$ indicating that
an energy gap between the solid orange curve and the dotted red curve appears 
at $q_x = f$.

\begin{figure}[htbp]
\centering
\includegraphics*[scale=0.6]{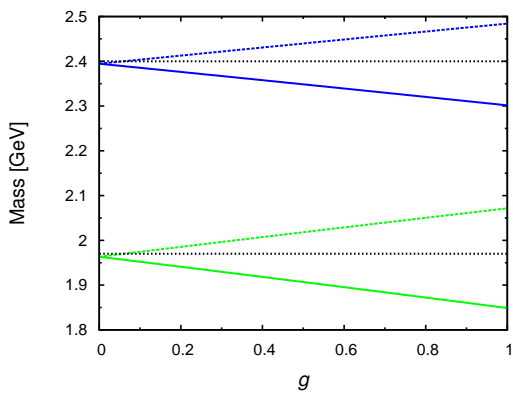}
\caption{(color online) $g$ dependence of the masses for $f=100$ MEV and $\tilde{\phi}=1$. 
The solid and dotted green curves show the masses corresponding to the solid and dotted green and red curves in Fig.~\ref{Dispersion}, while the blue curves to the blue and orange curves in Fig.~\ref{Dispersion}.  Two black dotted lines show the masses of $\bar{D}$ and $\bar{D}_0^\ast$ mesons in vacuum for comparison.}
\label{MassVSG}
\end{figure}

In order to see the mixing effect among spin-0 and spin-1 states clearly, we plot $g$ dependence of the masses for $f=100$ MeV and $\tilde{\phi}=1$ in Fig.~\ref{MassVSG}.
We note that, in Fig.~\ref{Dispersion}, two colored curves coincide at $q_x=k_\perp^2=0$, so that each curve in Fig.~\ref{MassVSG} shows the value of two degenerate masses:
the blue dotted curve corresponds to the blue and orange dotted curves in Fig.~\ref{Dispersion}, the blue solid curve to the blue and orange solid curves, the green dotted to the green and red dotted curves, and the green solid to the green and red solid curves.
Each of dotted curves coincide with one of the solid curves with the same color at $g=0$,
which shows that four modes have a degenerate mass at $g=0$.
From this figure we can see that, as $g$ increases, i.e., the mixing among ``$\bar{D}$ mesons'' becomes strong, the splitting between the masses of two modes drawn by the same color linearly increases proportionally to $gf$.

\subsection{Effects of mass differences}
\label{sec:MassDifference}

\begin{figure*}[t]
  \begin{center}
    \begin{tabular}{c}

  \subfigure[\ ($\bar{D}_u$, $\bar{D}_{0\, u}^{*}$)]{
      \begin{minipage}{0.48\hsize}
        \begin{center}
         \includegraphics*[scale=0.6]{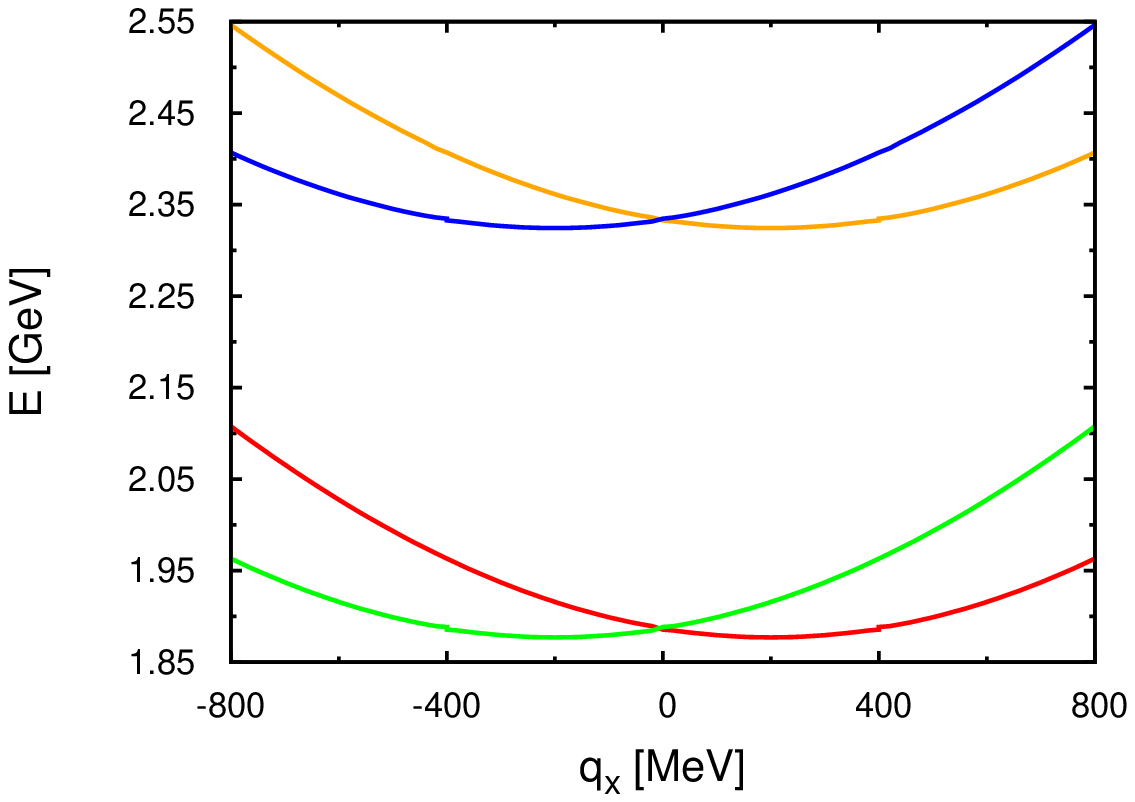}
        \end{center}
      \end{minipage}
}
  \subfigure[\ ($\bar{D}_u^{*1}$, $\bar{D}_{1\, u}^1$)]{
      \begin{minipage}{0.48\hsize}
        \begin{center}
          \includegraphics*[scale=0.6]{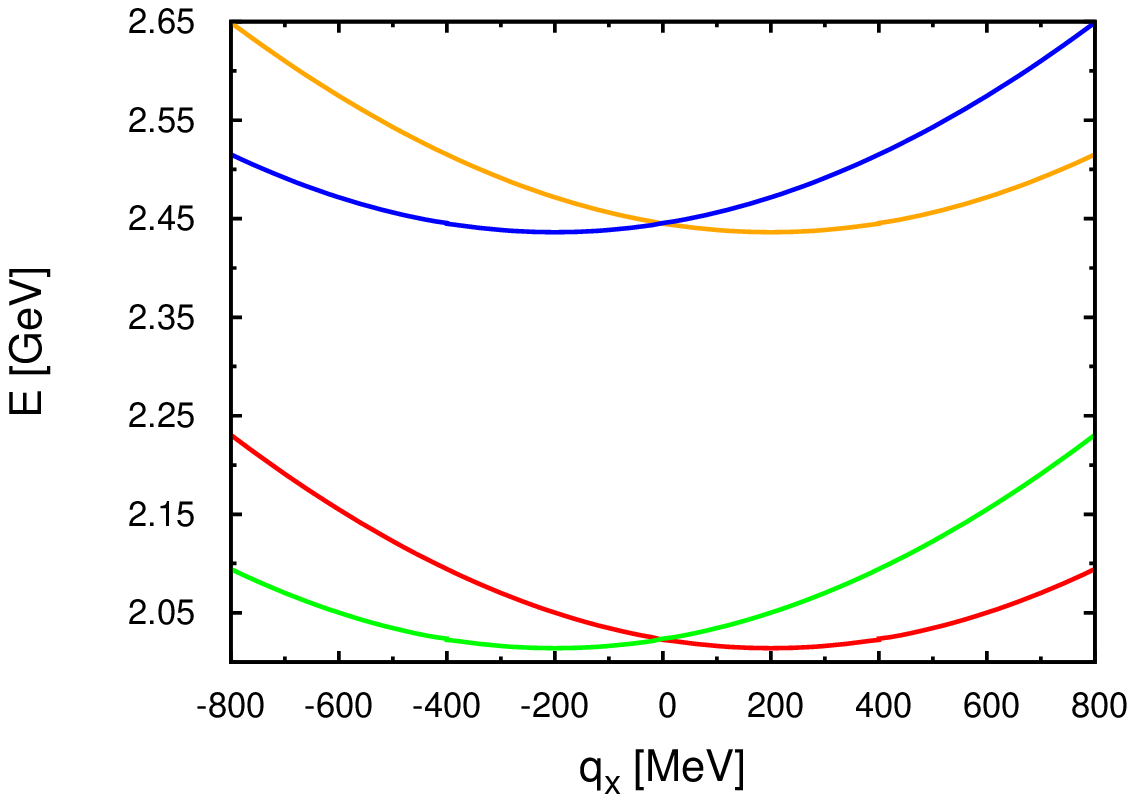} 
        \end{center}
      \end{minipage}
}
    \end{tabular}
 \caption{(color online) Dispersion relations 
for (a) ($\bar{D}_u$, $\bar{D}_{0\, u}^{*}$) and (b) ($\bar{D}_u^{*1}$, $\bar{D}_{1\, u}^1$) with $k_{\perp}^2=0$, $g=0$ and $f=200$ MeV,
when the mass difference between $\bar{D}_u$ and $\bar{D}_{0\, u}^{*}$ as well as that between $\bar{D}_u^{*1}$ and $\bar{D}_{1\, u}^1$ is included.
The blue and green curves show the dispersion relations for medium modified modes, while the orange and red curves are for the collective modes which have the negative group velocity in the small $q_x$ region.}
\label{MassDifference}
  \end{center}
\end{figure*}

In the real world, heavy quark symmetry is violated by $O(\Lambda_{\rm QCD}/m)$ correction, then the masses of $\bar{D}_u$ and $\bar{D}^{*1}_u$ states as well as those of $\bar{D}_{0\, u}^*$ and $\bar{D}_{1\, u}^1$ states are not identical.  
Here we study the spectra including the effect of mass differences as $m_{\bar{D}} =1869$ MeV, $m_{\bar{D}^*}=2010$ MeV, $m_{\bar{D}_0^*}=2318$ MeV, $m_{\bar{D}_1}=2427$ MeV. 
We show the dispersion relations 
for ($\bar{D}_u$, $\bar{D}_{0\, u}^{*}$) and ($\bar{D}_u^{*1}$, $\bar{D}_{1\, u}^1$) with $k_{\perp}^2=0$, $g=0$ and $f=200$ MeV in Fig.~\ref{MassDifference}. 
This shows that each meson has two modes: one has a positive velocity at $q_x=0$ and another has a negative velocity.
It should be noted that there are energy gaps when the momentum $q_x$ is  
the integral multiple of the wave number $f$, i.e., $q_x = n f$ with $n = 1, 2, \ldots$.
Since these gaps are at most of order $1$\,MeV, so that it is hard to see in the figure. 
We also note that  
each curve is expected to split into two curves for non-zero $g$ and sixteen modes 
appear 
for ($\bar{D}_u$, $\bar{D}_{0\, u}^{*}$, $\bar{D}_u^{*1}$, $\bar{D}_{1\, u}^1$).

\section{A summary and discussions}
\label{sec:DiscussionAndSummary}

In this paper, we studied the dispersion relations for ``$\bar{D}$ mesons'' in the dual chiral density wave (DCDW) phase where the chiral symmetry is spontaneously broken by an inhomogeneous chiral condensate.
Because the chiral condensate in the DCDW phase was expressed by a periodic function in a space coordinate, we employed the Bloch's theorem to get the dispersion relations. 
 We started with a relativistic Lagrangian based on the chiral partner structure where the pionic interactions of several ``$\bar{D}$ mesons'' were  
related with each other by the heavy quark symmetry.
The dispersion relations were calculated by solving the EoMs for ``$\bar{D}$ mesons'' in the presence of the background provided by the DCDW.  
Because of the $Z_2$ symmetry of the DCDW pointed 
in Ref.~\cite{Suenaga:2014dia}, 
dispersion relations for the ``charged $\bar{D}$ mesons'' were exactly the same as those for the ``neutral $\bar{D}$ mesons''. 
Furthermore, 
according to the EoMs~(\ref{EoMs}), the ``neutral $\bar{D}$ mesons'' were separated into two sectors: ($\bar{D}_u$, $\bar{D}^{\ast 1}_u$, $\bar{D}_{0\,u}^\ast$, $\bar{D}_{1\,u}^{1}$) mixed among themselves, while they did not mix with any of ($\bar{D}_u^{\ast 2}$, $\bar{D}^{\ast 3}_u$, $\bar{D}_{1\,u}^{2}$, $\bar{D}_{1\,u}^{3}$). 
Since the heavy quark symmetry existed for $E/M \to1$, 
we took into account only ($\bar{D}_u$, $\bar{D}_u^{*1}$, $\bar{D}_{0\, u}^*$, $\bar{D}_{1\, u}^1$) states. 

In Sec.~\ref{sec:MHMGWithoutG},
 we showed the resultant dispersion relations for $g=0$. In this case ($\bar{D}_u$, $\bar{D}^{\ast 1}_u$, $\bar{D}_{0\,u}^\ast$, $\bar{D}_{1\,u}^{1}$) states were further separated into two sectors: ($\bar{D}_u$, $\bar{D}_{0\, u}^*$) and ($\bar{D}_u^{*1}$, $\bar{D}_{1\, u}^1$), and EoMs for these two sectors were identical. This agreement
was understood by 
the $Z_2$ symmetry of flipping of the spin of the Brown Muck. 
Our results showed the existence of four modes
 in ($\bar{D}_u$, $\bar{D}_{0\, u}^*$) sector.
We should notice that two of them have a positive velocity at zero-momentum limit, while the other two have a negative velocity.
Furthermore, the 
minimum energy was realized at $q_x=-f$ or $q_x=f$.
Energy gaps did not occur although we employed the Bloch's theorem, which presumably reflects  
the chiral partner structure and the heavy quark symmetry. 
The  magnitude of chiral condensate $\phi$ could be measured by the energy difference of collective modes at given $q_x$. It was crucial that the potential form in the space-time coordinate was ${\rm sin}(2fx)$ and ${\rm cos}(2fx)$, and $\bar{D}$ and $\bar{D}_0^*$ mesons were related by chiral partner to each other, to get the eigenvalues~(\ref{Eigenvalues}) and corresponding eigenvectors~(\ref{EigenvectorMS}) with~(\ref{Eigenvectors}).

Next, we considered $g\neq0$ case in Sec.~\ref{sec:MHMG}. 
Structures of the dispersion relations were similar to the ones for $g=0$ except that the spin-0 sector ($\bar{D}_u$, $\bar{D}_{0\, u}^*$) and the spin-1 sector ($\bar{D}_u^{*1}$, $\bar{D}_{1\, u}^1$) were mixed to provide non-degenerate dispersion relations.
This mixing meant the $Z_2$ symmetry of the spin of Brown Muck disappeared. 
We finally took into account the effect of mass differences between $\bar{D}_u$ and $\bar{D}_u^{*1}$ states, $\bar{D}_{0\, u}^*$ and $\bar{D}_{1\, u}^1$ states.
The features of the spectra were similar to ones with heavy quark symmetry but there were tiny gaps of order $1$\,MeV at $q_x = n f$ with $n = 1, 2, \ldots$ reflecting the Bloch's treatment.

To summarize, our result shows that 
all the modes have 
a group velocity opposite to the momentum in the small momentum region and the energy is
minimized at non-zero momentum.
Furthermore,
the magnitude of momentum which realizes the minimum energy 
is equal to the wave number of the density wave.
This indicates that we can use  mesons including the heavy quark 
to probe the properties of high density medium.

We started with 
the effective Lagrangian~(\ref{StartingLagrangian}) originated from the Lagrangian~(\ref{LagrangianOfHMET}), where we assumed that only $g_1$ terms in~(\ref{LagrangianInTermsOfGH}) are dominant and set $g_2\sim g_4$ to be zero.
And we did not take into account terms including higher derivatives of chiral field $M$.
Although the inclusion of the terms with higher derivatives will change the dispersion relation, we expect that the feature of having the velocity opposite to the momemntum in the low momentrum region is not changed.
Furthermore, 
 the existence of Brillouin zone does not change so that we can extract the information of the wave number of the density wave $f$ by measuring the periodicity of dispersion relations for $\bar{D}$ mesons.
 We leave these studies for future publications. 

In Ref.~\cite{Suenaga:2014sga},
we considered the masses of ``$\bar{D}$ mesons'' 
in the Skyrme matter 
constructed by putting the Skyrmion on the face-centered cubic (FCC) crystal.
There, we took
the space average of classical solution of pion to obtain the masses of ``$\bar{D}$ mesons'',
and showed that the masses of $D$ ($J^P=0^-$) and $D_0^\ast$ ($0^+$) as well as those of $D^\ast$ ($1^-$) and $D_1$ ($1^+$) degenerated with each other.
This
is justified when the size of primitive cell of the Skyrme crystal $L$ is small: $(\frac{2\pi}{L})^2\gg \frac{\tilde{\phi}m\Delta_m}{2}$.

In the present analysis, we assumed that the $\sigma$- and $\pi$-type condensates were simply the cosine and sine functions of a space coordinate.
We can straightforwardly generalize the present analysis for more complicated structures for the condensate such as those in the Real Kink Soliton (RKS) phase~\cite{Abuki:2011pf}, where only the $\sigma$-type condensate exists and is described by a general 
periodic function.
In such a case, we expect that there appear many modes.

In this work we studied the dispersion relations for ``$\bar{D}$ mesons'' and showed the new modes which did not exist in the vacuum. 
For obtaining the 
knowledge about the pole residues of these modes, 
it is interesting 
to study also the spectral functions for these modes. 
We leave this to the future publications.

\acknowledgments

This work was supported in part by the JSPS Grant-in-Aid for Scientific Research (C) No.~24540266.



\appendix
\section{Heavy Meson Effective Theory with Chiral Partner Structure}
\label{sec:HMET}

 We start by an effective Lagrangian
of heavy-light mesons with chiral partner structure 
based
on the $SU(2)_h$ heavy quark spin symmetry, $SU(2)_L \times SU(2)_R$ chiral symmetry and parity 
invariance:
\begin{eqnarray}
{\cal L} &=& {\rm tr}[H_L(iv\cdot\partial)\bar{H}_L]+{\rm tr}[H_R(iv\cdot\partial)\bar{H}_R] \nonumber\\
&& +\frac{\Delta_m}{2f_{\pi}}{\rm tr}[H_LM\bar{H}_R + H_RM^{\dagger}\bar{H}_L] \nonumber\\
&& +i\frac{g_{1}}{2f_{\pi}}{\rm tr}[H_R\gamma_5\gamma^{\mu}\partial_{\mu}M^{\dagger}\bar{H}_L-H_L\gamma_5\gamma^{\mu}\partial_{\mu}M\bar{H}_R] \nonumber\\
&& + i\frac{g_{2}}{2f_{\pi}}{\rm tr}[H_L\gamma_5\Slash{\partial}MM^{\dagger}\bar{H}_L-H_R\gamma_5\Slash{\partial}M^{\dagger}M\bar{H}_R] \nonumber\\
&& +\frac{g_{3}}{2f_{\pi}}{\rm tr}[H_R\gamma^{\mu}\partial_{\mu}M^{\dagger}\bar{H}_L+H_L\gamma^{\mu}\partial_{\mu}M\bar{H}_R] \nonumber\\
&& +i\frac{g_{4}}{2f_{\pi}}{\rm tr}[H_L\Slash{\partial}MM^{\dagger}\bar{H}_L+H_R\Slash{\partial}M^{\dagger}M\bar{H}_R] \nonumber\\
&& + O(\partial^2M)\ , \label{HMETLagrangian}
\end{eqnarray}
where $\Delta_m$ is the mass difference between $G$ and $H$ doublet in the vacuum, $g_1$, $g_2$, $g_3$ and $g_4$ are the real coupling constants for the pionic interaction, $H_L$ and $H_R$ are heavy-light meson fields and $\bar{H}_{L(R)} = \gamma^0H^{\dagger}_{L(R)}\gamma^0$. $M$ is chiral field $M=\sigma+i\pi^a\tau^a$ where $\sigma$ and $\pi^a$ are scalar and pseudo-scalar fields respectively, and $\tau^a$ is the Pauli matrix.
$H_L$, $H_R$ and $M$ transform under the $SU(2)_L\times SU(2)_R$ chiral symmetry as
\begin{eqnarray}
H_{L(R)} \to H_{L(R)}g_{L(R)}^{\dagger} \ ,\ \  M \to g_L Mg_R^{\dagger}\ .
\end{eqnarray}
The parity eigenstates of heavy-light mesons are introduced as
\begin{eqnarray}
H_L &=& \frac{1}{\sqrt{2}}(G+iH\gamma_5) \nonumber\\
H_R &=& \frac{1}{\sqrt{2}}(G-iH\gamma_5)\ ,
\end{eqnarray}
where $H$ ($G$) is the negative (positive) parity state:
$H$ contains $(D,D^*)=(0^-,1^-)$ mesons and $G$ contains $(D_0^*,D_1)=(0^+,1^+)$ mesons. These are parametrized as
\begin{eqnarray}
H &=& \frac{1+\Slash{v}}{2}[i\gamma_5D+\Slash{D}^*] \nonumber\\
G &=& \frac{1+\Slash{v}}{2}[D_0^*-iD_1\gamma_5]\ , \label{PhysicalStates}
\end{eqnarray}
where $v^{\mu}$ is the velocity of $D$ mesons. In terms of $G$ doublet and $H$ doublet, the Lagrangian~(\ref{HMETLagrangian}) is written as
\begin{widetext}
\begin{eqnarray}
{\cal L} &=& {\rm tr}[G(iv\cdot\partial)\bar{G}]-{\rm tr}[H(iv\cdot\partial)\bar{H}] \nonumber\\
&& +\frac{\Delta_m}{4f_{\pi}}{\rm tr}[G(M+M^{\dagger})\bar{G}+H(M+M^{\dagger})\bar{H}-iG(M-M^{\dagger})\gamma_5\bar{H}+iH(M-M^{\dagger})\gamma_5\bar{G}] \nonumber\\
&&  +i\frac{g_{1}}{4f_{\pi}}{\rm tr}[G\gamma_5(\Slash{\partial}M^{\dagger}-\Slash{\partial}M)\bar{G}-H\gamma_5(\Slash{\partial}M^{\dagger}-\Slash{\partial}M)\bar{H} - iG(\Slash{\partial}M^{\dagger}+\Slash{\partial}M)\bar{H}-iH(\Slash{\partial}M^{\dagger}+\Slash{\partial}M)\bar{G}] \nonumber\\
&& +i\frac{g_{2}}{4f_{\pi}}{\rm tr}[G\gamma_5(\Slash{\partial}MM^{\dagger}-\Slash{\partial}M^{\dagger}M)\bar{G} +H\gamma_5(\Slash{\partial}MM^{\dagger}-\Slash{\partial}M^{\dagger}M)\bar{H}-iG(\Slash{\partial}MM^{\dagger}+\Slash{\partial}M^{\dagger}M)\bar{H} +iH(\Slash{\partial}MM^{\dagger}+\Slash{\partial}M^{\dagger}M)\bar{G}] \nonumber\\
&& +\frac{g_{3}}{4f_{\pi}}{\rm tr}[G(\Slash{\partial}M^{\dagger}+\Slash{\partial}M)\bar{G}-H(\Slash{\partial}M^{\dagger}+\Slash{\partial}M)\bar{H} - iG\gamma_5(\Slash{\partial}M^{\dagger}-\Slash{\partial}M)\bar{H}-iH\gamma_5(\Slash{\partial}M^{\dagger}-\Slash{\partial}M)\bar{G}] \nonumber\\
&& +i\frac{g_{4}}{4f_{\pi}}{\rm tr}[G(\Slash{\partial}MM^{\dagger}+\Slash{\partial}M^{\dagger}M)\bar{G} +H(\Slash{\partial}MM^{\dagger}+\Slash{\partial}M^{\dagger}M)\bar{H}-iG\gamma_5(\Slash{\partial}MM^{\dagger}-\Slash{\partial}M^{\dagger}M)\bar{H} +iH\gamma_5(\Slash{\partial}MM^{\dagger}-\Slash{\partial}M^{\dagger}M)\bar{G}] \ .\nonumber\\ \label{LagrangianInTermsOfGH}
\end{eqnarray}
\end{widetext}
This Lagrangian contains four terms including $\partial M$.
Here, we assume that $g_1$-term is dominant among these four terms, and we take $g_2=g_3=g_4=0$. 
Then using parametrization~(\ref{PhysicalStates}), the Lagrangian~(\ref{LagrangianInTermsOfGH}) can be written in terms of $D, D^*, D_0^*$ and $D_1$ mesons. 

Note that $D(\sim Q\bar{q})$ mesons contain anti-light quarks which provide annihilation process in the nuclear matter. 
Then,  in the present analysis, 
we consider $\bar{D}(\sim\bar{Q}q)$ mesons which
are defined as the charge conjugation of $D$ mesons, so that an effective Lagrangian of $\bar{D}$ mesons is obtained by taking charge conjugation of the above Lagrangian for $D$ mesons as
\begin{widetext}
\begin{eqnarray}
{\cal L} &=&  2\bar{D}(iv\cdot\partial)\bar{D}^{\dagger}-2\bar{D}_{\mu}^*(iv\cdot\partial)\bar{D}^{*\dagger\mu} + 2\bar{D}_0^*(iv\cdot\partial)\bar{D}_0^{*\dagger}-2\bar{D}_{1\mu}(iv\cdot\partial)\bar{D}_1^{\dagger\mu}  \nonumber\\
&& +  \frac{\Delta_m}{2f_{\pi}}[\bar{D}_0^*(M+M^{\dagger})\bar{D}_0^{*\dagger}-\bar{D}_{1\mu}(M+M^{\dagger})\bar{D}_1^{\dagger\mu}-\bar{D}(M+M^{\dagger})\bar{D}^{\dagger}+\bar{D}_{\mu}^*(M+M^{\dagger})\bar{D}^{*\dagger\mu}] \nonumber\\
 && +\frac{\Delta_m}{2f_{\pi}}[\bar{D}_0^*(M-M^{\dagger})\bar{D}^{\dagger}-\bar{D}_{1\mu}(M-M^{\dagger})\bar{D}^{*\dagger\mu}-\bar{D}(M-M^{\dagger})\bar{D}_0^{*\dagger}+\bar{D}_{\mu}^{*\dagger}(M-M^{\dagger})\bar{D}_1^{\dagger\mu}]\nonumber\\ 
&& -\frac{g_1}{2f_{\pi}}[\bar{D}_1^{\mu}(\partial_{\mu}M^{\dagger}-\partial_{\mu}M)\bar{D}_0^{*\dagger}-\bar{D}_0^{*}(\partial_{\mu}M^{\dagger}-\partial_{\mu}M)\bar{D}_1^{\dagger\mu}-\epsilon^{\mu\nu\rho\sigma}\bar{D}_{1\mu}(\partial_{\nu}M^{\dagger}-\partial_{\nu}M)\bar{D}_{1\rho}^{\dagger}v_{\sigma}] \nonumber\\
&& +\frac{g_1}{2f_{\pi}}[\bar{D}^{*\mu}(\partial_{\mu}M^{\dagger}-\partial_{\mu}M)\bar{D}^{\dagger}-\bar{D}(\partial_{\mu}M^{\dagger}-\partial_{\mu}M)\bar{D}^{*\dagger\mu}-\epsilon^{\mu\nu\rho\sigma}\bar{D}_{\mu}^*(\partial_{\nu}M^{\dagger}-\partial_{\nu}M)\bar{D}_{\rho}^{*\dagger}v_{\sigma}] \nonumber\\
&& +\frac{g_1}{2f_{\pi}}[\bar{D}_1^{\mu}(\partial_{\mu}M^{\dagger}+\partial_{\mu}M)\bar{D}^{\dagger}+\bar{D}(\partial_{\mu}M^{\dagger}+\partial_{\mu}M)\bar{D}_1^{\dagger\mu}]\nonumber\\
& & -\frac{g_1}{2f_{\pi}}[\bar{D}_0^*(\partial_{\mu}M^{\dagger}+\partial_{\mu}M)\bar{D}^{*\dagger\mu}+\bar{D}^{*\mu}(\partial_{\mu}M^{\dagger}+\partial_{\mu}M)\bar{D}_0^{*\dagger}]\nonumber\\
& & -\frac{g_1}{2f_{\pi}}[\epsilon^{\mu\nu\rho\sigma}\bar{D}_{1\nu}(\partial_{\rho}M^{\dagger}+\partial_{\rho}M)\bar{D}_{\mu}^{*\dagger}v_{\sigma}+\epsilon^{\mu\nu\rho\sigma}\bar{D}_{\mu}^*(\partial_{\rho}M^{\dagger}+\partial_{\rho}M)\bar{D}_{1\nu}^{\dagger}v_{\sigma}] \ . \label{LagrangianOfHMET}
\end{eqnarray}
\end{widetext}


\section{Bloch's Theorem}
\label{sec:BlochTheorem}
Let us consider one dimensional problem of quantum mechanics in a periodic potential with the period $a$ in a matter of size $L$. 
The wave function for a particle in the potential is expanded as
\begin{eqnarray}
\psi(x) = \sum_qC_q{\rm e}^{iqx}\ ,
\end{eqnarray}
where the wave number $q$ satisfies the boundary condition $q=2n\pi/L$. The periodic potential $V(x)$ with period $a$ can be expanded as
\begin{eqnarray}
V(x) = \sum_{K'}V_{K'}{\rm e}^{iK'x}\ ,
\end{eqnarray}
where $K'$ is the reciprocal lattice vector given by the integral multiple of $K=2\pi/a$.
Then the Schr\'odinger equation
\begin{eqnarray}
\left(-\frac{1}{2m}\frac{\partial^2}{\partial x^2}+V(x)\right)\psi(x)=\epsilon\psi(x)
\end{eqnarray}
leads to 
\begin{eqnarray}
\sum_q\left\{(\epsilon-\epsilon_{q})C_{q}-\sum_{K'}V_{K'}C_{k-K'}\right\}{\rm e}^{iqx}= 0\ \nonumber\\
\end{eqnarray}
where $\epsilon_q$ is the energy of the free particle with momentum $q$: $\epsilon_q = q^2/2m$.
From this, we get so-called the central equation in the periodic potential:
\begin{eqnarray}
(\epsilon-\epsilon_q)C_q-\sum_{K'}V_{K'}C_{q-K'} = 0 \ . \label{CentralEquation}
\end{eqnarray}
Here we define $q=k-\tilde{K}$ in such a way that $k$ lies in the first Brillouin zone as $-K/2 < k < K/2$. Then, this equation is rewritten into
\begin{eqnarray}
(\epsilon-\epsilon_{k-\tilde{K}})C_{k-\tilde{K}}-\sum_{K'}V_{K'-\tilde{K}}C_{k-K'} = 0 \ .
\end{eqnarray}
This equation is obtained with the expansion of wave function $\psi(x)$ as
\begin{eqnarray}
\psi(x) &=&\sum_k  \sum_{K'} C_{k-K'}{\rm e}^{i(k-K')x} \nonumber\\
&\equiv& \sum_{k} \psi_k(x) \ ,
\end{eqnarray}
where we have defined 
\begin{eqnarray}
\psi_k(x) = \sum_{K'}C_{k-K'}{\rm e}^{i(k-K')x}\ . \label{WaveFunctionOfk}
\end{eqnarray}
This wave function is often written as
\begin{eqnarray}
\psi_k(x) = {\rm e}^{ikx}\sum_{K'}C_{k-K'}{\rm e}^{-K'x} \equiv {\rm e}^{ikx}u_k(x)\ ,\label{BlochFunction}
\end{eqnarray}
and easily confirm the function $u_k(x)$ is periodic with period $L$:
\begin{eqnarray}
u_k(x+L) &\equiv& \sum_{K'}C_{k-K'}{\rm e}^{-iK'(x+L)} = \sum_{K'}C_{k-K'}{\rm e}^{-iK'x} \nonumber\\
 &=& u_k(x)\ .\label{PeriodicFunction}
 \end{eqnarray}
Furthermore, we can find $\psi_k(x)$ satisfies a relation
\begin{eqnarray}
\psi_{k}(x) = \psi_{k\pm K}(x) = \psi_{k\pm2K}(x) = \cdots
\end{eqnarray}
from Eq.~(\ref{WaveFunctionOfk}), then energy eigenvalues $\epsilon$ in periodic potentials are also periodic in the momentum space:
\begin{eqnarray}
\epsilon(k) = \epsilon(k\pm K) = \epsilon(k\pm2K)=\cdots\ . \label{EigenvaluesPeriodicity}
\end{eqnarray}

The form of the wave function~(\ref{BlochFunction}) is called Bloch function, and what we mentioned in this Appendix is called the Bloch's theorem.


\section{Derivation of Eqs.~(\ref{Eigenvalues})-(\ref{Dispersion12}) }
\label{sec:ExtendedZone}
In this appendix, we explain how we obtain Eqs.~(\ref{Eigenvalues})-(\ref{Dispersion12}), 
and why we have only four curves in Fig.~\ref{DispersionWithoutG}, 
 although we can expect infinite number of modes in the periodic potentials as the band structure of electron in a metal. 
 
We first solve the EoMs (\ref{EoMsMatrix}) with taking into account only three Brillouin zones around the first zone and determine the eigenvalues and corresponding eigenvectors. By defining 
\begin{eqnarray}
e_n^2 = (k_x+nK)^2
\end{eqnarray}
and
\begin{eqnarray}
v = \frac{m\Delta_m}{2}\tilde{\phi}\ ,
\end{eqnarray}
Eq.~(\ref{EoMsMatrix}) is reduced to the following eigenvalue equation for the $6\times 6$ matrix in this truncation: 
\begin{widetext}
\begin{eqnarray}
\left(
\begin{array}{cccccc}
-\bar{E}^2+e_{-1}^2 & 0 & -v & -v & 0 & 0 \\
0 &  -\bar{E}^2+e_{-1}^2 & v & v & 0 & 0 \\
-v & v & -\bar{E}^2+ e_{0}^2 & 0 & -v & -v \\
-v & v & 0 &  -\bar{E}^2+ e_{0}^2 &  v & v \\
0 & 0 & -v & v & -\bar{E}^2+e_1^2 & 0  \\
0 & 0 & -v & v & 0 & -\bar{E}^2+e_1^2 \\
\end{array}
\right) \left(
\begin{array}{c}
C_{k_x-K}^{\bar{D}_u} \\
C_{k_x-K}^{\bar{D}_{0u}^*} \\
C_{k_x}^{\bar{D}_u} \\
C_{k_x}^{\bar{D}_{0u}^*} \\
C_{k_x+K}^{\bar{D}_u} \\
C_{k_x+K}^{\bar{D}_{0u}^*} \\
\end{array}
\right) = 0\ .  \label{EoMj1}
\end{eqnarray}
Six eigenvalues and corresponding eigenvectors up to the normalization factor for this eigenvalue equation are summarized as follows: 
\begin{enumerate}
 \renewcommand{\theenumii}{\arabic{enumii}}
 \renewcommand{\labelenumi}{(F\theenumi)}
\item For $\bar{E}^2=(k_x-jK)^2$ 
\begin{eqnarray}
\left(1,1,0,0,0,0\right)^T\ , \nonumber
\end{eqnarray} 
\label{enum:j11}
\item For $\bar{E}^2=(k_x+jK)^2$
\begin{eqnarray}
\left(0,0,0,0,-1,1\right)^T\ , \nonumber
\end{eqnarray}
\label{enum:j12}
\item For $\bar{E}^2 = \frac{1}{2}\left[e_0^2+e_{-1}^2\pm\sqrt{(e_0^2-e_{-1}^2)^2+16v^2}\right] $ 
\begin{eqnarray}
\left(\frac{4v}{e_{-1}^2-e_0^2\mp\sqrt{(e_{-1}^2-e_0^2)^2+16v^2}},-\frac{4v}{e_{-1}^2-e_0^2\mp\sqrt{(e_{-1}^2-e_0^2)^2+16v^2}}, 1, 1, 0, 0\right)^T \ , \nonumber
\end{eqnarray}
\label{enum:j13}
\item For $\bar{E}^2 =  \frac{1}{2}\left[e_1^2+e_{0}^2\pm\sqrt{(e_1^2-e_{0}^2)^2+16v^2}\right]$
\begin{eqnarray}
 \left(0, 0, \frac{4v}{e_{0}^2-e_1^2\mp\sqrt{(e_{0}^2-e_1^2)^2+16v^2}},-\frac{4v}{e_{0}^2-e_1^2\mp\sqrt{(e_{0}^2-e_1^2)^2+16v^2}}, 1, 1\right)^T \nonumber
\end{eqnarray}
 \label{enum:j14}
\end{enumerate}
\end{widetext}
(double sign correspondence). We can notice that the eigenvector in~(F\ref{enum:j13}) has non-vanishing values only for $C_{k_x-K}^{\bar{D}_u}$, $C_{k_x-K}^{\bar{D}_{0\, u}^*}$, $C_{k_x}^{\bar{D}_u}$ and $C_{k_x}^{\bar{D}_{0\, u}^*}$, while~(F\ref{enum:j14}) has only for $C_{k_x}^{\bar{D}_u}$, $C_{k_x}^{\bar{D}_{0\, u}^*}$, $C_{k_x+K}^{\bar{D}_u}$ and $C_{k_x+K}^{\bar{D}_{0\, u}^*}$. We would like to stress that~(\ref{EoMj1}) is solved analytically because of the mixing between $\bar{D}_u$ and $\bar{D}_{0\, u}^*$ inside a given Brillouin zone in addition to mixings with the nearest Brillouin zones, and the particular form of potentials. These features are provided by the chiral partner structure, 
and the Fourier transformation of the potential written in terms of $\cos(2fx)$ and $\sin(2fx)$. 

\begin{figure*}[t]
  \begin{center}
    \begin{tabular}{c}

      \begin{minipage}{0.33\hsize}
        \begin{center}
         \includegraphics*[scale=0.5]{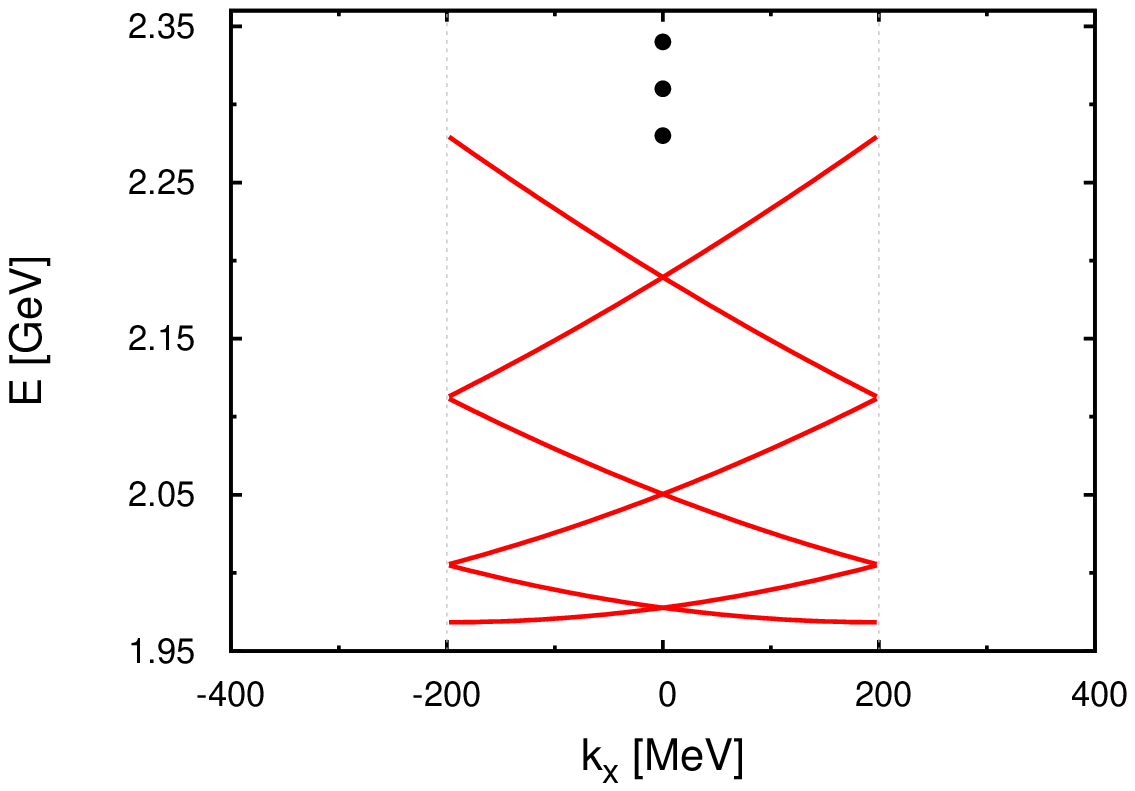}
          \hspace{0.5cm} (a) 
          First step
        \end{center}
      \end{minipage}

      \begin{minipage}{0.33\hsize}
        \begin{center}
          \includegraphics*[scale=0.5]{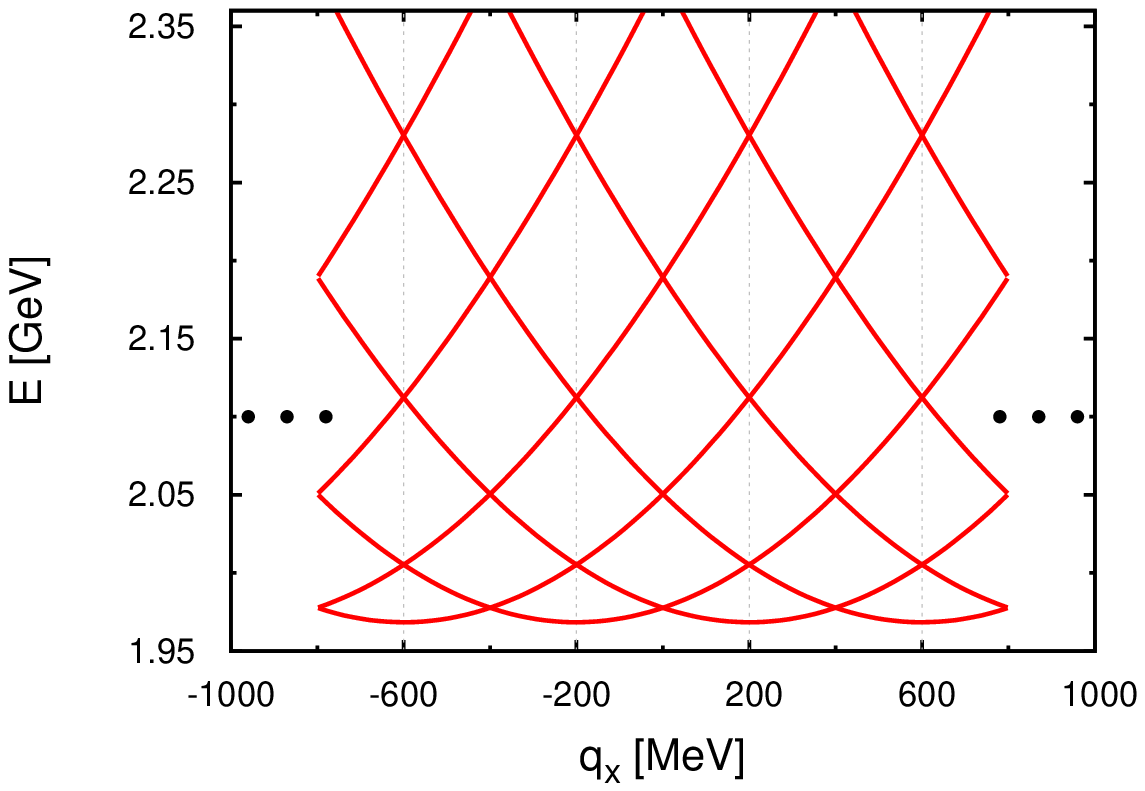}
          \hspace{0.5cm} (b) 
          Second step
        \end{center}      
      \end{minipage}

      \begin{minipage}{0.33\hsize}
        \begin{center}
          \includegraphics*[scale=0.5]{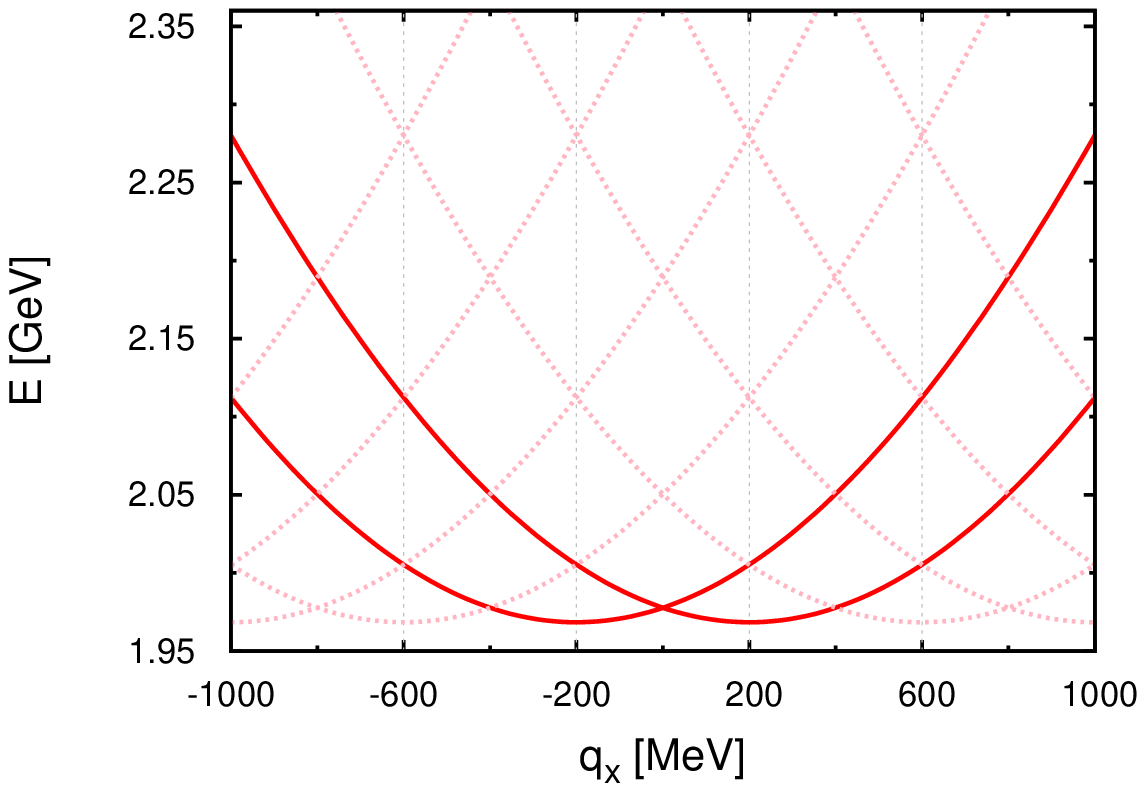}
          \hspace{0.5cm} (c) 
          Third step
        \end{center}
      \end{minipage}

    \end{tabular}
   \caption{(color online)Three steps to derive Eqs.~(\ref{Dispersion11}) and (\ref{Dispersion12}). Here, the dispersion relation for $k_{\perp}^2=0$, $\tilde{\phi}=1$ and $f=200$ MeV is used as an example.  
Three steps are provided to get these results, and we show the procedure for lower two modes plotted in Fig.~\ref{DispersionWithoutG}. See the text detail is stated in Appendix~\ref{sec:ExtendedZone}. }
\label{ExtendedZones}
  \end{center}
\end{figure*}

Next, we solve EoM~(\ref{EoMsMatrix}) generally, and show how to obtain Eqs.~(\ref{Eigenvalues})-(\ref{Eigenvectors}).
In order to get Eq.~(\ref{Eigenvalues}), we take into account $2j+1$ Brillouin zones around the first zone, i.e., $C_{k_x-jK},\cdots, C_{k_x+jK}$, so that we diagonalize the $2(2j+1)\times2(2j+1)$ matrix~\footnote{
Note that an additional factor $2$ arises since we consider $\bar{D}_u$ and $\bar{D}_{0\, u}^*$ system here, and $C_{q}=(C_{q}^{\bar{D}_u}, C_q^{\bar{D}_{0\, u}^*})^T$.
}  in Eq.~(\ref{EoMsMatrix}). The 
eigenvalues are obtained analytically as
\begin{eqnarray}
 \bar{E}^2 &=&(k_x-jK )^2\nonumber\\
 &,& \frac{1}{2}\left[e_{-j+1}^2+e_{-j}^2 \pm \sqrt{(e_{-j+1}^2-e_{-j}^2)^2+16v^2}\right]  \nonumber\\
 &,& \frac{1}{2}\left[e_{-j+2}^2+e_{-j+1}^2\pm \sqrt{(e_{-j+2}^2-e_{-j+1}^2)^2+16v^2} \right] \nonumber\\
 &,&\ \ \ \ \ \ \ \ \ \ \vdots\ \ \ \ \ \vdots\ \ \ \ \ \vdots\nonumber\\
  &,& \frac{1}{2}\left[e_{j-1}^2+e_{j-2}^2 \pm\sqrt{(e_{j-1}^2-e_{j-2}^2)^2+16v^2}\right]  \nonumber\\
 &,& \frac{1}{2}\left[e_j^2+e_{j-1}^2\pm \sqrt{(e_{j}^2-e_{j-1}^2)^2+16v^2}\right]  \nonumber\\
 &,& (k_x+jK)^2\ .
 \end{eqnarray}
These eigenvalues are collectively expressed by using a label $n$ which runs from $-j$ to $j-1$ as
\begin{eqnarray}
\bar{E}^2 &=& \frac{1}{2}\left[e_{n+1}^2+e_n^2 \pm \sqrt{(e_{n+1}^2-e_n^2)^2+16v^2} \right]
 \label{EigenvaluesWithN}
 \end{eqnarray}
except $\bar{E}^2=(k_x-jK)^2$ and $\bar{E}^2=(k_x+jK)^2$. 
We can easily see that eigenvalues in Eq.~(\ref{EigenvaluesWithN}) is nothing but the eigenvalues in Eq.~(\ref{Eigenvalues}).   
The corresponding eigenvectors generally take $2(2j+1)$ components column as
\begin{equation}
\begin{pmatrix}
C_{k_x-jK} \\
\vdots \\
C_{k_x-K} \\
C_{k_x}\\
C_{k_x+K} \\
\vdots \\
C_{k_x+jK} \\
\end{pmatrix}
\end{equation}
\\
with
\begin{equation}
C_{k_x + n K} = 
\begin{pmatrix}
C_{k_x + n K}^{\bar{D}_u} \\
C_{k_x + n K}^{\bar{D}_{0\, u}^*} \\
\end{pmatrix}
\ , \quad ( n = - j , \ldots, j) \ .
\end{equation}
Up to the normalization factor, the components of the eigenvectors are obtained analytically as follows: 
 \begin{widetext}
\begin{enumerate}
 \renewcommand{\theenumii}{\arabic{enumii}}
 \renewcommand{\labelenumi}{(G\theenumi)}
\item For $\bar{E}^2=(k_x-jK)^2$ 
\begin{eqnarray}
C_{k_x-jK} =\left(
\begin{array}{c}
C_{k_x-jK}^{\bar{D}_u}\\
C_{k_x-jK}^{\bar{D}_{0\, u}^*}\\
\end{array}
\right) = \left(
\begin{array}{c}
1\\
1\\
\end{array}
\right)\ ,\ \   
{\rm and\ the\ other\ components\ are\ zero.}  \nonumber
\end{eqnarray}
\label{enum:Case1}
\item For $\bar{E}^2=(k_x+jK)^2$ 
\begin{eqnarray}
C_{k_x+jK} =\left(
\begin{array}{c}
C_{k_x+jK}^{\bar{D}_u}\\
C_{k_x+jK}^{\bar{D}_{0\, u}^*}\\
\end{array}
\right) = \left(
\begin{array}{c}
-1\\
1\\
\end{array}
\right)\ ,\ \    {\rm and\ the\ other\ components \ are\ zero.}  \nonumber
\end{eqnarray}
\label{enum:Case2}

\item For $\bar{E}^2 = \frac{1}{2}\left[e^2_{n+1}+e_{n}^2\pm\sqrt{(e_{n+1}^2-e_{n}^2)^2+16v^2}\right]$ \ \  ($n=-j,\cdots, j-1$)
 \begin{eqnarray} 
&& C_{k_x+nK} = \left(
 \begin{array}{c}
 C_{k_x+nK}^{\bar{D}_u} \\
 C_{k_x+nK}^{\bar{D}_{0\, u}^*}\\
 \end{array}
 \right)
= \left (
\begin{array}{c}\displaystyle
\frac{4v}{e_n^2-e_{n+1}^2\mp\sqrt{(e_n^2-e_{n+1}^2)^2+16v^2}}\\\displaystyle
-\frac{4v}{e_n^2-e_{n+1}^2\mp\sqrt{(e_n^2-e_{n+1}^2)^2+16v^2}}\\
\end{array}
\right)   \ , \ \ 
 C_{k_x+(n+1)K} =\left(
\begin{array}{c}
C_{k_x+(n+1)K}^{\bar{D}_u}\\
C_{k_x+(n+1)K}^{\bar{D}_{0\, u}^*}\\
\end{array}
\right) = \left(
\begin{array}{c}
1\\
1\\
\end{array}
\right)\nonumber\\\nonumber\\
&&    {\rm and\ the\ other\ coefficients \ are\ zero.}  \nonumber
\end{eqnarray}
 \label{enum:Case3}
\end{enumerate}
\end{widetext}
Note that the double-sign in $C_{k_x+nK}$ in~(G\ref{enum:Case3}) corresponds to the double-sign
of corresponding eigenvalues. Note also that eigenvectors in~(G\ref{enum:Case3}) is identical to~(\ref{Eigenvectors}). When we take $j \to \infty$ limit, the eigenvalues $\bar{E}^2=(k_x-jK)^2$ and $\bar{E}^2=(k_x+jK)^2$ have large values compared with the energy region in which we are interested in, then we can neglect these two eigenvalues and corresponding eigenvectors~(G\ref{enum:Case1}) and~(G\ref{enum:Case2}). So we conclude that the eigenvalues for EoMs~(\ref{EoMsMatrix}) are Eq.~(\ref{Eigenvalues}) and corresponding eigenvectors are of the form~(\ref{EigenvectorMS}) with~(\ref{Eigenvectors}).



Finally, we explain how to get Eqs.~(\ref{Dispersion11}) and (\ref{Dispersion12}), and plot Fig.~\ref{DispersionWithoutG}.
Here we show how to derive lower two modes (red and green curves) in Fig.~\ref{DispersionWithoutG}. We need three steps which are pictorially shown in Fig.~\ref{ExtendedZones}, where the dispersion relation for $k_{\perp}^2=0$, $\tilde{\phi}=1$ and $f=200$ MeV is used as an example. 
First, we get the eigenvalues~(\ref{EigenvaluesWithN}) with  taking the $j\to \infty$ limit as we explained above.
The crystal momentum $k_x$ is restricted in the first Brillouin zone $-f\leq k_x\leq f$, and a  number of curves characterized by Eq.~(\ref{EigenvaluesWithN})
actually appear when we make a plot only in this interval as shown in Fig.~\ref{ExtendedZones} (a).  
Second, using the relation~(\ref{EigenvaluesPeriodicity}), we extend these modes out of the first zone, i.e., we rewrite them in terms of momentum $q_x$. 
This dispersion relation is periodic with period $K=2f$ as shown in Fig~\ref{ExtendedZones} (b). 
Third, from the eigenvalues and corresponding eigenvectors~(G\ref{enum:Case3}) with $j\to\infty$ limit, we find only neighboring components $C_{k_x+nK}$ and $C_{k_x+(n+1)K}$ have non-vanishing values for each $n$-th eigenvalues in~(\ref{EigenvaluesWithN}). For example, when $q_x$ lies in $K/2 < q_x < 3 K/2$, there are only two eigenvalues for which the component $C_{k_x+K}$ is non-vanishing as shown in Fig~\ref{ExtendedZones} (c).
In this way, we draw lower two (red and green) curves in Fig.~\ref{DispersionWithoutG}. 

Blue and orange curves in Fig.~\ref{DispersionWithoutG} are obtained by the same way and we can finally get the four curves in Fig.~\ref{DispersionWithoutG}.



\end{document}